\documentclass[prd,superscriptaddress,unsortedaddress,twocolumn,showpacs,preprintnumbers,amsmath,amssymb]{revtex4}
\usepackage[dvipdfmx]{graphicx}
\usepackage{amsmath,amssymb,times}
\usepackage{ulem}
\usepackage{braket}

\def\fun#1#2{\lower3.6pt\vbox{\baselineskip0pt\lineskip.9pt
\ialign{$\mathsurround=0pt#1\hfil##\hfil$\crcr#2\crcr\sim\crcr}}}

\newcommand{\beq}{\begin{equation}}
\newcommand{\eeq}{\end{equation}}
\newcommand{\bea}{\begin{eqnarray}}
\newcommand{\eea}{\end{eqnarray}}

\newcommand{\la}{\left\langle}

\DeclareSymbolFont{boldletters}{OML}{cmm} {b}{it}
\DeclareSymbolFontAlphabet{\mathbit}{boldletters}
\DeclareMathSymbol{\alpha}{\mathalpha}{letters}{"0B}
\DeclareMathSymbol{\beta}{\mathalpha}{letters}{"0C}
\DeclareMathSymbol{\gamma}{\mathalpha}{letters}{"0D}
\DeclareMathSymbol{\delta}{\mathalpha}{letters}{"0E}
\DeclareMathSymbol{\epsilon}{\mathalpha}{letters}{"0F}
\DeclareMathSymbol{\zeta}{\mathalpha}{letters}{"10}
\DeclareMathSymbol{\eta}{\mathalpha}{letters}{"11}
\DeclareMathSymbol{\theta}{\mathalpha}{letters}{"12}
\DeclareMathSymbol{\iota}{\mathalpha}{letters}{"13}
\DeclareMathSymbol{\kappa}{\mathalpha}{letters}{"14}
\DeclareMathSymbol{\lambda}{\mathalpha}{letters}{"15}
\DeclareMathSymbol{\mu}{\mathalpha}{letters}{"16}
\DeclareMathSymbol{\nu}{\mathalpha}{letters}{"17}
\DeclareMathSymbol{\xi}{\mathalpha}{letters}{"18}
\DeclareMathSymbol{\pi}{\mathalpha}{letters}{"19}
\DeclareMathSymbol{\rho}{\mathalpha}{letters}{"1A}
\DeclareMathSymbol{\sigma}{\mathalpha}{letters}{"1B}
\DeclareMathSymbol{\tau}{\mathalpha}{letters}{"1C}
\DeclareMathSymbol{\upsilon}{\mathalpha}{letters}{"1D}
\DeclareMathSymbol{\phi}{\mathalpha}{letters}{"1E}
\DeclareMathSymbol{\chi}{\mathalpha}{letters}{"1F}
\DeclareMathSymbol{\psi}{\mathalpha}{letters}{"20}
\DeclareMathSymbol{\omega}{\mathalpha}{letters}{"21}
\DeclareMathSymbol{\varepsilon}{\mathalpha}{letters}{"22}
\DeclareMathSymbol{\vartheta}{\mathalpha}{letters}{"23}
\DeclareMathSymbol{\varpi}{\mathalpha}{letters}{"24}
\DeclareMathSymbol{\varrho}{\mathalpha}{letters}{"25}
\DeclareMathSymbol{\varsigma}{\mathalpha}{letters}{"26}
\DeclareMathSymbol{\varphi}{\mathalpha}{letters}{"27}
\DeclareMathSymbol{\Gamma}{\mathalpha}{letters}{"00}
\DeclareMathSymbol{\Delta}{\mathalpha}{letters}{"01}
\DeclareMathSymbol{\Theta}{\mathalpha}{letters}{"02}
\DeclareMathSymbol{\Lambda}{\mathalpha}{letters}{"03}
\DeclareMathSymbol{\Xi}{\mathalpha}{letters}{"04}
\DeclareMathSymbol{\Pi}{\mathalpha}{letters}{"05}
\DeclareMathSymbol{\Sigma}{\mathalpha}{letters}{"06}
\DeclareMathSymbol{\Upsilon}{\mathalpha}{letters}{"07}
\DeclareMathSymbol{\Phi}{\mathalpha}{letters}{"08}
\DeclareMathSymbol{\Psi}{\mathalpha}{letters}{"09}
\DeclareMathSymbol{\Omega}{\mathalpha}{letters}{"0A}

\def\la{\mathrel{\mathpalette\fun <}}
\def\ga{\mathrel{\mathpalette\fun >}}
\def\fun#1#2{\lower3.6pt\vbox{\baselineskip0pt\lineskip.9pt
\ialign{$\mathsurround=0pt#1\hfil##\hfil$\crcr#2\crcr\sim\crcr}}}

\begin{document}
\title{Model prediction for temperature dependence of meson pole masses \\ 
from lattice QCD results on meson screening masses}

\author{Masahiro Ishii}
\email[]{ishii@phys.kyushu-u.ac.jp}
\affiliation{Department of Physics, Graduate School of Sciences, Kyushu University,
             Fukuoka 819-0395, Japan}

\author{Hiroaki Kouno}
\email[]{kounoh@cc.saga-u.ac.jp}
\affiliation{Department of Physics, Saga University,
             Saga 840-8502, Japan}  

\author{Masanobu Yahiro}
\email[]{yahiro@phys.kyushu-u.ac.jp}
\affiliation{Department of Physics, Graduate School of Sciences, Kyushu University,
             Fukuoka 819-0395, Japan}             

\date{\today}

\begin{abstract} 
We propose a practical effective model by introducing 
temperature ($T$) dependence to the coupling strengths of  
four-quark and six-quark Kobayashi-Maskawa-'t Hooft interactions in 
the 2+1 flavor Polyakov-loop extended Nambu--Jona-Lasinio model. 
The $T$ dependence is determined from LQCD data on 
the renormalized chiral condensate around the pseudocritical temperature $T_c^{\chi}$  of 
chiral crossover and 
the screening-mass difference between $\pi$ and $a_0$ mesons 
in $T > 1.1T_c^\chi$ where 
only the $U(1)_{\rm A}$-symmetry breaking survives. 
The model well reproduces LQCD data 
on screening masses $M_{\xi}^{\rm scr}(T)$ 
for both scalar and pseudoscalar mesons, 
particularly in $T \ga T_c^{\chi}$. 
Using this effective model, we predict meson pole masses 
$M_{\xi}^{\rm pole}(T)$ for scalar and pseudoscalar mesons. 
For $\eta'$ meson, the prediction is consistent with 
the experimental value at finite $T$ measured in heavy-ion collisions. 
We point out that the relation 
$M_{\xi}^{\rm scr}(T)-M_{\xi}^{\rm pole}(T) \approx 
M_{\xi'}^{\rm scr}(T)-M_{\xi'}^{\rm pole}(T)$ 
is pretty good when $\xi$ and $\xi'$ are scalar mesons, 
and show that the relation 
$M_{\xi}^{\rm scr}(T)/M_{\xi'}^{\rm scr}(T) \approx 
M_{\xi}^{\rm pole}(T)/M_{\xi'}^{\rm pole}(T)$ is 
well satisfied within 20\% error when $\xi$ 
and $\xi'$ are pseudoscalar mesons and also when $\xi$ 
and $\xi'$ are scalar mesons. 
\end{abstract}

\maketitle

%%%%%%%%%%%%%%%%%%%%%%%%%%%%%%%%%%%%%%%%%%%%%%%%%%%%%%%%%%%%%%%%%%%%%%%%%%%
%%%%%  Introduction 
%%%%%%%%%%%%%%%%%%%%%%%%%%%%%%%%%%%%%%%%%%%%%%%%%%%%%%%%%%%%%%%%%%%%%%%%%%%
\section{Introduction}
\label{Introduction}
Meson masses are fundamental quantities characterizing 
hadron properties. 
Temperature ($T$) dependence of 
meson masses plays an important role in understanding 
properties of hot-QCD matter, for example, in 
determining reaction rates of hadron-hadron
collisions and dilepton production. 
In fact, $T$ dependence of $\eta'$ and vector mesons was recently measured in heavy-ion collisions~\cite{Csorgo,STAR:2014}.

For later convenience, we call the meson mass ``meson pole mass''
in order to distinguish it from ``meson screening mass''. 
Meson pole and screening masses, $M_{\xi}^{\rm pole}$ and 
$M_{\xi}^{\rm scr}$, of $\xi$ meson are defined by the inverse of 
the exponential decay of the mesonic correlation function $\zeta_{\xi
\xi}(\tau,{\bf x})$ 
in its temporal $\tau$- and spatial ${\bf x}$-directions, respectively. 
As seen from the definition, 
$M_{\xi}^{\rm pole}$ is 
experimentally measurable, but $M_{\xi}^{\rm scr}$ is not. 
In first-principle lattice QCD (LQCD) simulations at finite $T$, meanwhile, 
$M_{\xi}^{\rm scr}(T)$ is often calculated instead of $M_{\xi}^{\rm pole}(T)$, 
since the spatial lattice size is larger than  the temporal one 
for finite $T$. 
The relation between $M_{\xi}^{\rm pole}(T)$ and $M_{\xi}^{\rm scr}(T)$ is 
not understood for finite $T$, although 
$M_{\xi}^{\rm pole}(0)=M_{\xi}^{\rm scr}(0)$ from the definition.

As already mentioned above, 
meson screening masses are relatively easier to calculate 
than meson pole masses in LQCD simulations at finite $T$, 
since the spatial lattice size is larger than  the temporal one; 
see Appendix \ref{Difficulty of meson pole mass calculations in LQCD} 
for further discussion on the difficulty of meson pole-mass calculations. 
In fact, $T$ dependence of light-meson screening masses was 
recently determined 
in a wide range $140 \la T \la 800$ MeV 
by 2+1 flavor LQCD simulations with 
improved (p4) staggered fermions~\cite{Cheng:2010fe}. 
Meson screening masses are thus 
available with LQCD simulations, although they are 
not experimentally measurable.

Meson screening masses are useful in investigating 
symmetric properties of hot-QCD matter. 
In principle, 
one can understand the chiral-symmetry restoration 
through  $T$ dependence of the mass difference 
$\Delta M_{\sigma,\pi}^{\rm scr}(T)=M_{\sigma}^{\rm scr}(T)-M_{\pi}^{\rm scr}(T)$ between chiral-partner mesons, say $\pi$ and $\sigma$ mesons, 
and can analyze 
``the effective $U(1)_{\rm A}$-symmetry restoration'' 
through $T$ dependence of  
$\Delta M_{a_0,\pi}^{\rm scr}(T)=M_{a_0}^{\rm scr}(T)-M_{\pi}^{\rm scr}(T)$ 
between $U(1)_{\rm A}$-partner mesons, say $\pi$ and $a_0$ mesons. 
In the operator level, $U(1)_{\rm A}$ symmetry is explicitly broken by 
$U(1)_{\rm A}$ anomaly for any $T$, but in the expectation-value level 
the symmetry is restored at high $T$ 
by the suppression of topologically-nontrivial 
gauge configurations responsible for $U(1)_{\rm A}$ anomaly. 
The restoration is called ``effective $U(1)_{\rm A}$-symmetry
restoration''. For the effective $U(1)_{\rm A}$-symmetry restoration, 
we cannot define the order parameter clearly, but 
we may consider the difference $\Delta M_{a_0,\pi}^{\rm scr}(T)$
as an  indicator of the restoration. 
At the present stage, however, $\Delta M_{a_0,\pi}^{\rm scr}(T)$ 
is available but $\Delta M_{\sigma,\pi}^{\rm scr}(T)$ is 
not in LQCD simulations of Ref.~\cite{Cheng:2010fe}, 
because difficult LQCD calculations 
with the quark-line disconnected diagrams are necessary for 
$M_{\sigma}^{\rm scr}(T)$. 
Parallel discussion may be possible for $\Delta M_{\sigma,\pi}^{\rm pole}(T)=M_{\sigma}^{\rm pole}(T)-M_{\pi}^{\rm pole}(T)$ and 
$\Delta M_{a_0,\pi}^{\rm pole}(T)
=M_{a_0}^{\rm pole}(T)-M_{\pi}^{\rm pole}(T)$ if the $T$ dependence is experimentally measured.

For the 2+1 flavor system composed of light u- and d-quark
with the same mass $m_l$ and s-quark with the mass $m_s$, 
the renormalized chiral condensate
\begin{eqnarray}
 \Delta_{l,s}(T)
  = \frac{\sigma_l(T) - \frac{m_l}{m_s}\sigma_s(T)}
              {\sigma_l(0) - \frac{m_l}{m_s}\sigma_s(0)} 
\label{eq:renormalized chiral condensate}
\end{eqnarray}
is commonly used as an order parameter (indicator) of the chiral-symmetry 
restoration~\cite{Cheng:2008,Borsanyi:2010bp,Bazavov:2011nk}, 
where $\sigma_l(T)$ ($\sigma_s(T)$) is the chiral condensate 
for light quarks (s-quark). The chiral-symmetry restoration is 
found to be crossover \cite{Aoki_etal}
and the pseudocritical temperature $T_c^{\chi}$ is determined 
to be $T_c^{\chi}=154\pm 9$ MeV~\cite{Borsanyi:2010bp,Bazavov:2011nk}.
$T$ dependence of $\Delta_{l,s}(T)$ is also obtainable with LQCD, but 
not directly measurable with experiments.

Physical quantities, $\Delta_{l,s}(T)$ and $M_{\xi}^{\rm scr}(T)$, 
are thus obtainable with LQCD simulations 
but not with experiments. 
On the contrary, $M_{\xi}^{\rm pole}(T)$ is experimentally measurable but 
hard to get with LQCD simulations. 
If we can predict $M_{\xi}^{\rm pole}(T)$ theoretically from LQCD data 
on $\Delta_{l,s}(T)$ and $M_{\xi}^{\rm scr}(T)$, 
this makes it possible to compare the $M_{\xi}^{\rm pole}(T)$ 
predicted from LQCD data 
with the corresponding experimental data directly. 
If experimental data are not available for 
$M_{\xi}^{\rm pole}(T)$ of interest, such a prediction may be helpful in 
experimental analyses.

As a complementary approach  to 
 LQCD simulations, 
one can consider effective models 
such as {the Nambu--Jona-Lasinio (NJL) model and }the the Polyakov-loop extended Nambu--Jona-Lasinio (PNJL) model
~\cite{KHKB,Meisinger,Dumitru,Fukushima1,Costa:2005,Ghos,Megias,Ratti1,Ciminale,Ratti2,Rossner,Hansen,Sasaki-C,Schaefer,Kashiwa1,Costa:2009,Ruivo:2012a}. 
As already mentioned above, $M_{\xi}^{\rm pole}(T)$  are not easy 
to calculate with LQCD simulations. 
In addition, if one is interested in physical quantities at 
finite quark chemical potential $\mu$, 
LQCD simulations face the well-known sign problem, so that LQCD results are 
concentrated on the $\mu/T \la 1$ region. For this reason, the phase diagram beyond the region has been 
discussed and predicted with effective models. 
In particular, the PNJL model has been applied for many phenomena, since it 
can treat both chiral and deconfinement transitions. Recently, 
a Polyakov-loop ($\Phi$) dependent four-quark vertex was introduced 
to control the 
correlation between the two transitions~\cite{Sakai_EPNJL,Sasaki_EPNJL}. 
The PNJL model with the entanglement ($\Phi$-dependent) 
four-quark interaction is called 
the entanglement-PNJL (EPNJL) model~\cite{Sakai_EPNJL,Sasaki_EPNJL}. 
The EPNJL model is 
quite successful in reproducing LQCD data in the imaginary $\mu$ region~\cite{D'Elia3,FP2010} and the real isospin chemical potential region~\cite{Kogut2} 
where LQCD is free from the sign problem. 

$T$ dependence of $M_{\xi}^{\rm pole}(T)$ for low-lying meson was studied
extensively with NJL-type effective
models~\cite{KHKB,Costa:2005,Hansen,Costa:2009,Ruivo:2012a,Ruivo:2012b}. In spite of the success of NJL-type models 
in reproducing meson pole masses at $T=0$, 
it was difficult to calculate $M_{\xi}^{\rm scr}(T)$ 
with NJL-type models. However, 
this problem was solved very lately by formulating 
the meson correlation function carefully 
in momentum space~\cite{Ishii:2013kaa}; 
see Sec. \ref{Meson screening masses} for the detail. 
For the 2+1 flavor system, NJL-type effective models 
usually consist of the scalar-type four-quark 
interaction responsible for the chiral-symmetry restoration and 
the Kobayashi-Maskawa-'t Hooft (KMT) determinant (six-quark) 
 interaction~\cite{KMK,tHooft} responsible for 
the effective $U(1)_{\rm A}$-symmetry restoration. 
In general, the coupling strength $G_{\rm D}$ 
of the KMT interaction 
is proportional to the $T$-dependent instanton density 
$dn_{\rm inst}(T)$~\cite{Pisarski-Yaffe-1980,Shuryak:1993ee}.
For high $T$, the instanton density $dn_{\rm inst}(T)$ is suppressed by the Debye-type screening~\cite{Pisarski-Yaffe-1980,Shuryak:1993ee}. 
This means that  $G_{\rm D}$ depends on $T$: $G_{\rm D}=G_{\rm D}(T)$. 
Very lately, $T$ dependence of $G_{\rm D}(T)$ was determined from LQCD data 
on $\Delta M_{a_0,\pi}^{\rm scr}(T)$ with 
the EPNJL model \cite{Ishii:2015ira}. 
The $G_{\rm D}(T)$ thus determined predicts that there is 
 a tricritical point of chiral phase transition 
in the southwest direction of the physical point 
on $m_l$--$m_s$ plane.  
The success of the EPNJL model in reproducing LQCD data is originated 
in the fact that the coupling strength $G_{\rm S}$ 
of the scalar-type four-quark 
interaction depends on $T$ through $\Phi$. 
The EPNJL model is thus essentially equal to 
the PNJL model with a $T$-dependent coupling strength 
$G_{\rm S}=G_{\rm S}(T)$, 
as far as the case of $\mu=0$ is concerned.

In this paper, we propose a practical effective model by introducing 
$T$-dependent coupling strengths, $G_{\rm S}(T)$ and $G_{\rm D}(T)$, 
to the 2+1 flavor PNJL model. 
$T$ dependence of $G_{\rm S}(T)$ is determined 
from LQCD data on $T_c^{\chi}$~\cite{Borsanyi:2010bp,Bazavov:2011nk} 
and 
$\Delta_{l,s}(T)$~\cite{Cheng:2008}, while 
$T$ dependence of $G_{\rm D}(T)$ is from LQCD data 
on $\Delta M_{a_0,\pi}^{\rm scr}(T)$~\cite{Cheng:2010fe} 
in $T > 1.1T_c^\chi=170$~MeV 
where only the $U(1)_{\rm A}$-symmetry breaking survives 
\cite{Bhattacharya:2014ara,Ishii:2015ira}. 
In $T > 1.04T_c^{\chi}=160$ MeV, 
this model reproduces LQCD data~\cite{Cheng:2010fe} on 
$M_{\xi}^{\rm scr}(T)$ for both pseudoscalar mesons 
$\pi,K,\eta_{\bar{s}s}$ and scalar mesons 
$a_0,\kappa,\sigma_{\bar{s}s}$. 
In $T < 1.04T_c^{\chi}=160$ MeV, the 
agreement between model results and LQCD data is good for 
pseudoscalar $\pi,K$ mesons and pretty good for scalar 
$a_0,\kappa,\sigma_{\bar{s}s}$ mesons. 
For $\eta_{\bar{s}s}$ meson, 
the model result overestimates 
LQCD data by about $10\% \sim 30\%$ 
in $T < 1.04T_c^{\chi}=160$ MeV, but the deviation 
becomes small rapidly as $T$ increases from $160$ MeV. 
The deviation may be related to the fact that 
the disconnected diagrams are neglected in LQCD calculations. 
This point is discussed.

Using this practical effective model, we predict meson pole masses 
$M_{\xi}^{\rm pole}(T)$ for pseudoscalar mesons 
$\pi,K,\eta,\eta'$ and scalar mesons 
$a_0,\kappa,\sigma,f_0$. 
For $\eta'$ meson, the prediction  is compared with 
the experimental value~\cite{Csorgo} at finite $T$ measured in heavy-ion collisions. 
We show that the relation 
$M_{\xi}^{\rm scr}(T)-M_{\xi}^{\rm pole}(T) \approx 
M_{\xi'}^{\rm scr}(T)-M_{\xi'}^{\rm pole}(T)$ 
is pretty good when $\xi$ and $\xi'$ are scalar mesons, 
and point out that the relation 
$M_{\xi}^{\rm scr}(T)/M_{\xi'}^{\rm scr}(T) \approx 
M_{\xi}^{\rm pole}(T)/M_{\xi'}^{\rm pole}(T)$ is 
well satisfied within 20\% error when $\xi$ 
and $\xi'$ are pseudoscalar mesons and also when $\xi$ 
and $\xi'$ are scalar mesons. 
These relations may be useful to estimate $M_{\xi}^{\rm pole}$ 
for lighter $\xi$-meson  
from $M_{\xi'}^{\rm pole}$ and $M_{\xi'}^{\rm scr}$ for heavier $\xi'$-meson  
that may be obtainable with 
state-of-art LQCD simulations.

In Sec. \ref{Formalism}, we explain the present model and 
show the methods of evaluating $M_{\xi}^{\rm pole}(T)$ 
and $M_{\xi}^{\rm scr}(T)$. 
Numerical results are shown in Sec. \ref{Numerical Results}. 
Section \ref{Summary} is devoted to a summary.

%%%%%%%%%%%%%%%%%%%%%%%%%%%%%%%%%%%%%%%%%%%%%%%%%%%%%%%%%%%%%%%%%%%%%%%%%%%
%%%%% PNJL model
%%%%%%%%%%%%%%%%%%%%%%%%%%%%%%%%%%%%%%%%%%%%%%%%%%%%%%%%%%%%%%%%%%%%%%%%%%%
\section{Formalism}
\label{Formalism}

\subsection{Model setting}
\label{Model setting}

We consider the 2+1 flavor PNJL model~\cite{Meisinger,Dumitru,Costa:2005,Fukushima1,Ghos,Megias,Ratti1,Ciminale,Ratti2,Rossner,Hansen,Sasaki-C,Schaefer,Kashiwa1,Costa:2009,Ruivo:2012a} and introduce $T$-dependent coupling strengths, $G_{\rm S}(T)$ and $G_{\rm D}(T)$, for four- and six-quark interactions. 
The Lagrangian density is 
\begin{align}
 {\cal L}  
=& {\bar \psi}(i \gamma_\nu D^\nu - {\hat m_0} )\psi  
  + G_{\rm S}(T) \sum_{a=0}^{8} 
    [({\bar \psi} \lambda_a \psi )^2 +({\bar \psi }i\gamma_5 \lambda_a \psi )^2] 
\nonumber\\
 &- G_{\rm D}(T) \Bigl[\det_{f,f'} {\bar \psi}_f (1+\gamma_5) \psi_{f'} 
           +\det_{f,f'} {\bar \psi}_f (1-\gamma_5) \psi_{f'} \Bigr]
\nonumber\\
&-{\cal U}(\Phi [A],{\bar \Phi} [A],T) ,  
\label{L}
\end{align} 
where the gauge field $A^\nu$ in $D^\nu=\partial^\nu + iA^\nu$ is assumed to be $A^\nu=g \delta^{\nu}_{0}(A^0)_a{t_a / 2}=-ig \delta^{\nu}_{0}(A_{4})_a t_a/2$ for the gauge coupling $g$. The $\lambda_a$ ($t_a$) are the Gell-Mann matrices in flavor (color) space, 
$\lambda _0 = \sqrt{2/3}~{\boldsymbol I}_{\rm F}$ for 
the unit matrix ${\boldsymbol I}_{\rm F}$ in flavor space, 
and $\det_{f,f'}$ stands for the determinant  in flavor space. 
In the 2+1 flavor system, 
the quark fields $\psi=(\psi_u,\psi_d,\psi_s)^T$ 
have current quark masses 
${\hat m_0}={\rm diag}(m_u,m_d,m_s)$ satisfying $m_s > m_l \equiv m_u=m_d$.

In the original version of the PNJL model, 
the coupling strength $G_{\rm D}$ of KMT six-quark interaction is constant, 
but it has been shown very recently 
in Ref. \cite{Ishii:2015ira} 
that $T$ dependence is necessary for $G_{\rm D}$ 
 to explain LQCD data on $T$ dependence of 
$\Delta M_{a_0,\pi}^{\rm scr}$, 
i.e., the $U(1)_{\rm A}$-symmetry restoration. 
The $T$-dependent strength $G_{\rm D}(T)$ thus determined is  
%%%%%%%%%%%%%%%%%%%%
 \begin{eqnarray}
  G_{\rm D}(T)
   =\left\{
     \begin{array}{ll}
 G_{\rm D}(0) & (T < T_1) \\
 G_{\rm D}(0) e^{-(T-T_1)^2/b_1^2}& (T \ge T_1) \\
     \end{array}
     \right..
 \label{T-dependent-K}
 \end{eqnarray}
%%%%%%%%%%%%%%%%%%%
$T$ dependence of Eq. \eqref{T-dependent-K} 
is consistent with that of the instanton density 
$dn_{\rm inst}(T)$~\cite{Pisarski-Yaffe-1980,Shuryak:1993ee} 
for high $T$.

It is very likely that $T$ dependence is necessary also for 
the coupling strength $G_{\rm S}$ of four-quark interaction. 
We then introduce a $T$-dependent coupling strength $G_{\rm S}(T)$ 
of the same function form as $G_{\rm D}(T)$:
%%%%%%%%%%%%%%%%%%%%
 \begin{eqnarray}
 G_{\rm S}(T)=\left\{ \begin{array}{ll}
 G_{\rm S}(0) & (T < T_2) \\
 G_{\rm S}(0) e^{-(T-T_2)^2/b_2^2}& (T \ge T_2) \\
 \end{array} \right.. 
 \label{T-dependent-G}
 \end{eqnarray}
 %%%%%%%%%%%%%%%%%%%
It is possible to determine the parameter set  $(T_1,b_1)$ from LQCD data on 
$\Delta M_{a_0,\pi}^{\rm scr}(T)$ and the set $(T_2,b_2)$ from LQCD data 
on $\Delta_{l,s}(T)$. 
The results of this parameter fitting 
will be shown in Sec. \ref{Parameter fitting}. 
The resultant values are tabulated 
in Table \ref{Model parameters in coupling strengths}. 
In our previous work~\cite{Ishii:2015ira}, we used 
the EPNJL model with a $T$-dependent KMT interaction of 
form \eqref{T-dependent-K}. 
The values of $T_1$ and $b_1$ are close to the present ones shown 
in Table \ref{Model parameters in coupling strengths}.

%%%%%%%%%%%%%%%%%%%%%%%%%%%%%%
%   Table *
%%%%%%%%%%%%%%%%%%%%%%%%%%%%%%
\begin{table}[h]
\begin{center}
\caption
{Model parameters in coupling strengths $G_{\rm S}(T)$ and $G_{\rm D}(T)$. 
 }

\begin{tabular}{lccccccc}
 \hline \hline
$T_1$~[{\rm MeV}] & $b_1$~[{\rm MeV}]  & $T_2$~[{\rm MeV}] & $b_2$~[{\rm MeV}]
\\
\hline
 121~&43.5~&131~&83.3\\
 \hline
\end{tabular}
 \label{Model parameters in coupling strengths}
\end{center}
\end{table}
%%%%%%%%%%%%%%%%%%%%%%%%%%%%%%

In the PNJL model, only the time component $A_4$ of gauge field $A_\mu$ 
is treated as a homogeneous and static background field. 
In the Polyakov gauge, the Polyakov loop $\Phi$ and its 
Hermitian conjugate ${\bar \Phi}$ are obtained  by  
%%%%%%%%%%%%%%%%%%%
\begin{align}
\Phi &= {1\over{3}}{\rm tr}_{\rm c}(L),
~~~~~{\bar \Phi} ={1\over{3}}{\rm tr}_{\rm c}({L^*})
\label{Polyakov}
\end{align}
%%%%%%%%%%%%%%%%%%%
with the Polyakov-loop operator 
\bea
L= \exp[i A_4/T]=\exp[i \times {\rm diag}(A_4^{11},A_4^{22},A_4^{33})/T] 
\eea
for real variables $A_4^{jj}$ 
satisfying $A_4^{11}+A_4^{22}+A_4^{33}=0$, 
where the symbol ${\rm tr}_{\rm c}$ denotes the trace in color space. 
For the case of  $\mu=0$ where $\Phi={\bar \Phi}$, 
one can set $A^{33}_4=0$ and determine the others as 
$A^{22}_4=-A^{11}_4={\rm cos}^{-1}[(3\Phi -1)/2]$. 

We take the logarithm-type Polyakov-loop potential of Ref.~\cite{Rossner} 
as $\mathcal{U}$ that is determined from  $T$ dependence of LQCD data in the
pure gauge limit. When the potential is applied to QCD with dynamical
quarks, the parameter $T_0$ included in $\mathcal{U}$ is used as
an adjustable parameter. In the present case, we take $T_0=180$ MeV so
that the PNJL model can reproduce 2+1 flavor LQCD data on $T$ dependence
of pion screening mass at $T \ga T_c^\chi=154\pm9$~MeV, where the value of 
$T_c^\chi$ is determined with LQCD simulations~\cite{Borsanyi:2010bp,Bazavov:2011nk}.

Making the mean field approximation (MFA) to Eq. \eqref{L}, one 
can obtain the linearized Lagrangian density 
%%%%%%%%%%%%%
\begin{eqnarray}
  {\cal L}^{\rm MFA}  
= {\bar \psi}S^{-1}\psi  - U_{\rm M} - {\cal U}(\Phi [A],{\bar \Phi} [A],T) , 
\label{linear-L}
\end{eqnarray}
where the quark propagator 
\bea
S=(i \gamma_\nu \partial^\nu - \gamma_0A^0 - \hat{M})^{-1} 
\eea
depends on the chiral condensates $\sigma_f = \langle \bar
\psi_f \psi_f\rangle$ ($f=u,d,s$) through 
the effective-mass matrix 
 $\hat{M}={\rm diag}(M_u,M_d,M_s)$ with 
\begin{eqnarray*}
M_u &=& m_u -4G_{\rm S}(T)\sigma_u +2G_{\rm D}(T)\sigma_d \sigma_s ,\\
M_d &=& m_d -4G_{\rm S}(T)\sigma_d +2G_{\rm D}(T)\sigma_s \sigma_u ,\\
M_s &=& m_s -4G_{\rm S}(T)\sigma_s +2G_{\rm D}(T)\sigma_u \sigma_d . 
\end{eqnarray*}
The mesonic potential $U_{\rm M}$ is defined by 
\begin{eqnarray*}
U_{\rm M}
= 2G_{\rm S}(T)(\sigma^2_u+\sigma^2_d+\sigma^2_s)  
-4G_{\rm D}(T) \sigma_u \sigma_d \sigma_s.
\end{eqnarray*}

Making the path integral over quark fields in the mean-field action,  
one can get   
the thermodynamic potential (per unit volume) 
\begin{align}
&\Omega= U_{\rm M}+{\cal U}-2 \sum_{f=u,d,s} \int \frac{d^3 {\bf p}}{(2\pi)^3}
   \Bigl[ 3 E_{{\bf p},f} \notag \\
&+ \frac{1}{\beta}
           \ln~ [1 + 3(\Phi+{\bar \Phi} e^{-\beta E_{{\bf p},f})} 
           e^{-\beta E_{{\bf p},f}}+ e^{-3\beta E_{{\bf p},f}}] \notag\\
&+ \frac{1}{\beta} 
           \ln~ [1 + 3({\bar \Phi}+{\Phi e^{-\beta E_{{\bf p},f}}}) 
              e^{-\beta E_{{\bf p},f}}+ e^{-3\beta E_{{\bf p},f}}]
	      \Bigl]
\label{PNJL-Omega}
\end{align}
with 
$E_{{\bf p},f}=\sqrt{{\bf p}^2+M^2_f}$ and $\beta = 1/T$. 
The mean-field variables 
($X=\sigma_l,\sigma_s,\Phi,\bar{\Phi}$) 
are determined by the stationary conditions 
\begin{equation}
\frac{\partial \Omega}{\partial X}=0,
\end{equation}
where isospin symmetry is assumed for the light-quark sector, 
i.e., $\sigma_l \equiv \sigma_u=\sigma_d$ and $M_l = M_u = M_d$.

On the right-hand side of Eq. (\ref{PNJL-Omega}), the first term (vacuum term) 
diverges. The three-dimensional (3d) momentum-cutoff regularization is often 
used to avoid the divergence. However, the regularization breaks Lorentz
invariance and thereby induces an unphysical oscillation in 
the spatial correlation function $\zeta_{\xi\xi}(0,{\bf x})$
\cite{Florkowski}.
In addition, the fundamental 
relation $M_{\xi}^{\rm pole}(0)=M_{\xi}^{\rm scr}(0)$ 
is not satisfied as a consequence of the Lorentz-symmetry breaking. We then use the Pauli-Villars (PV) regularization~\cite{Florkowski,PV}. 
This PV regularization has a parameter $\Lambda$ with mass dimension; 
see Sec. \ref{Meson screening masses} for further explanation.

%%%%%%%%%%%%%%%%%%%%%%%%%%%%%%
%   Table *
%%%%%%%%%%%%%%%%%%%%%%%%%%%%%%
\begin{table}[h]
\begin{center}
\caption
{Model parameters determined from physical quantities at vacuum. 
Set (A) is the realistic parameter set that is  
determined from experimental or empirical values at vacuum. 
In set (B), $m_l$ and $m_s$ are slightly changed from set (A) so as to 
become consistent with the lattice setting 
($m_l/m_s=1/10$ and $M_{\pi}^{\rm pole}(0)=176$ MeV) 
of LQCD simulations of Ref.~\cite{Cheng:2010fe,Cheng:2008}. 
}
\begin{tabular}{lcccccccc}
 \hline \hline
&$m_l$~[{\rm MeV}] & $m_s$~[{\rm MeV}]  & $G_{\rm S}(0)\Lambda^2$
& $G_{\rm D}(0)\Lambda^5$~ & $\Lambda$~[{\rm MeV}]
\\
\hline
 set (A) :& 8~&191~&2.72~&~40.4~&~660\\
 set (B) :& 13~&130~&2.72~&~40.4~&~660\\
\hline
\end{tabular}
 \label{Other model parameters}
\end{center}
\end{table}
%%%%%%%%%%%%%%%%%%%%%%%%%%%%%%

Eventually, the present model has five parameters 
$(m_l, m_s, G_{\rm S}(0), G_{\rm D}(0), \Lambda)$ 
in addition to $T_0$, $(T_1,b_1)$ and $(T_2,b_2)$. 
The five parameters can be determined 
from experimental or empirical values at vacuum. 
The determination of the five parameters should be made before the determination of 
$T_0$, $(T_1,b_1)$ and $(T_2,b_2)$. 
We first assume $m_l=8$~MeV and then determine the values of 
$(m_s, G_{\rm S}(0), G_{\rm D}(0), \Lambda)$ so as to 
reproduce experimental data, 
$f_\pi=92.4$~MeV, $M_{\pi}^{\rm pole}=138$~MeV, 
$M_{K}^{\rm pole}=495$~MeV and $M_{\eta'}^{\rm pole}=958$~MeV, 
where $f_\pi$ is the pion decay constant.  
The resulting parameter values are shown 
as set (A) in Table \ref{Other model parameters}. When we compare model results with LQCD data, we refit the values of $m_l$ and $m_s$ 
so as to become consistent with the lattice 
setting. This parameter set is refered to as set (B) in this paper; see
Sec. \ref{Model tuning for LQCD-data analyses} for the detail.

Table \ref{input+output} shows physical quantities at vacuum 
calculated with the parameter set (A) of Table \ref{Other model parameters} and the corresponding experimental or empirical values. 
Numbers with asterisk are inputs of the present parameter fitting.
The parameter set (A) reproduces available 
experimental data reasonably well. 
In addition, the results of set (A) are close to those of the parameter set in 
Ref. \cite{KHKB} for meson pole masses for $\eta,a_0,\kappa,\sigma,f_0$, 
the mixing angle $\theta_\eta$ between $\eta_0$ and $\eta_8$ states, 
the mixing angle $\theta_\sigma$ between $\sigma_0$ and $\sigma_8$ states, 
the effective s-quark mass $M_s$, and the kaon decay constant $f_K$.  

%%%%%%%%%%%%%%%%%%%%%%%%%%%%%%
%   Table *
%%%%%%%%%%%%%%%%%%%%%%%%%%%%%%
\begin{table}[h]
\begin{center}
\caption
{
Physical quantities at vacuum calculated with the parameter set (A) of  
Table \ref{Other model parameters} and 
the corresponding experimental or empirical values. Numbers with asterisk 
are inputs of the present parameter fitting. 
Experimental data are taken from Refs.~\cite{PDG2014,CBR}. 
The effective light-quark mass $M_l \approx 336$~MeV is 
estimated from experimental data on baryon magnetic
 moments~\cite{CBR}. Since we impose the isospin symmetry, we estimate 
experimental values of averaged pion and kaon masses as $M_\pi\equiv (M_{\pi^0}^{\rm exp} + M_{\pi^+}^{\rm exp} +
 M_{\pi^-}^{\rm exp})/3=(134.97 + 2\times 139.57)/3=138.0$~{\rm MeV} and
 $M_K\equiv (M_{K^0}^{\rm exp} + M_{\bar{K}^0}^{\rm exp} + M_{K^+}^{\rm exp} + M_{K^-}^{\rm exp})/4=(2\times 497.61 + 2\times
 493.68)/4=495.6$~{\rm MeV}. Experimental data on the 
decay constants $f_\pi$ and $f_K$ are taken for charged pion and kaon. 
 } 
\begin{tabular}{lcccccccc}
 \hline \hline
~& $M_\pi$~[{\rm MeV}]  & $M_K$~[{\rm MeV}] & $M_{\eta'}$~[{\rm MeV}] & $f_\pi$~[{\rm MeV}]&$f_K$~[{\rm MeV}]
\\
\hline
Cal.~&$138^\ast$~&$495^\ast$~&~$958^\ast$~&~$92.4^\ast$ & 96.2
                 \\
Exp.~&$138.0$~&$495.6$~&~$957.8$~&~$92.2$ &110.5
                 \\
 \hline \hline
~&$M_\eta$~[{\rm MeV}]  & $M_{a_0}$~[{\rm MeV}] &
             $M_{\kappa}$~[{\rm MeV}] &
                     $M_{\sigma}$~[{\rm MeV}] &$M_{f_0}$~[{\rm MeV}]
                     \\
\hline
Cal.& 487~&813~&~1016~&~674~&~1185~
                     \\
Exp.& 547.8~&980$\pm20$~&~800~&~400$\sim$550~&~980$\pm$20~
                     \\
 \hline \hline
&$\theta_{\eta}$ & $\theta_\sigma$ & $M_{l}$~[{\rm MeV}] & $M_{s}$~[{\rm MeV}] &
                     \\
\hline
Cal.&$-7.40^\circ$~&~$17.6^\circ$~&~336~&544~&\\ 
Exp.&$-11.4^\circ$~&~--~&~336~&--&\\ 
 \hline
\end{tabular}
  \label{input+output}
\end{center}
\end{table}
%%%%%%%%%%%%%%%%%%%%%%%%%%%%%%

%%%%%%%%%%%%%%%%%%%%%%%%%%%%%%%%%%%%%%%%%%%%%%%%%%%%%%%%%%%%%%%%%%%%%%%%%%%
%%%%% Correlation function
%%%%%%%%%%%%%%%%%%%%%%%%%%%%%%%%%%%%%%%%%%%%%%%%%%%%%%%%%%%%%%%%%%%%%%%%%%%
\subsection{Meson pole masses}
We consider pseudoscalar mesons ($\xi=\pi,K,\eta,\eta'$) and scalar ones 
($\xi=a_0,\kappa,\sigma,f_0$), 
and recapitulate the formalism of Refs.~\cite{KHKB}. 
The current operator for $\xi$ meson is expressed by  
\beq
  J_{\xi}(x) = \bar \psi(x)\Gamma_\xi \psi(x)
             - \langle\bar \psi(x) \Gamma_\xi \psi(x)\rangle  
\label{source}
\eeq
with $\Gamma_\xi= {\boldsymbol I}_{\rm C} \otimes \Gamma_{\rm D}
\otimes\Gamma_{\rm F}$, where ${\boldsymbol I}_{\rm C}$ is the unit matrix 
in color space. 
The matrix $\Gamma_{\rm D}$ in Dirac space is 
$\Gamma_{\rm D}={\boldsymbol I}_{\rm D}$ for the scalar channel and
$\Gamma_{\rm D}=i\gamma_5$ for the pseudoscalar channel, 
where ${\boldsymbol I}_{\rm D}$ is the unit matrix 
in Dirac space. The matrix $\Gamma_{\rm F}$ in flavor space is 
\begin{eqnarray}
 \Gamma_{\rm F}
  =
\left\{
\begin{array}{cccc}
 \lambda_3                         & {\rm for} & \pi,~~a_0        \\
(\lambda_4\pm i\lambda_5)/\sqrt{2} & {\rm for} & K,~~\kappa        \\
\lambda_{\rm s}                    & {\rm for} & \eta_{\bar{s}s},~~\sigma_{\bar{s}s}\\
\lambda_{\rm ns}                   & {\rm for} & \eta_{\bar{l}l},~~\sigma_{\bar{l}l}
\end{array}
\right. , 
\end{eqnarray}
where $\lambda_{\rm ns}={\rm diag}(1,1,0)$ and 
$\lambda_{\rm s}={\rm diag}(0,0,\sqrt{2})$.

Mesons $\eta$ and $\eta'$ ($\sigma$ and $f_0$) are described as mixed states 
of $\eta_{\bar{s}s}$ and $\eta_{\bar{l}l}$ ($\sigma_{\bar{s}s}$ and $\sigma_{\bar{l}l}$) states: Namely, 
\begin{eqnarray}
 \left(
\begin{array}{c}
 \eta'\\
 \eta
\end{array}
\right)
=
 O(\theta_\eta^{ls})
 \left(
\begin{array}{c}
 \eta_{\bar{s}s}\\
 \eta_{\bar{l}l}
\end{array}
              \right) 
 ,~~
  \left(
\begin{array}{c}
 f_0\\
 \sigma
\end{array}
\right)
=
O(\theta^{ls}_{\sigma})
\left(
\begin{array}{c}
 \sigma_{\bar{s}s}\\
 \sigma_{\bar{l}l}
\end{array}
                             \right)
\nonumber
\\
\label{angle}
\end{eqnarray}
with the orthogonal matrix $O(\theta)$
\begin{equation}
 O(\theta) =
 \left(
\begin{array}{cc}
 \cos{\theta}&\sin{\theta}\\
 -\sin{\theta}&\cos{\theta}
\end{array}
\right),
\end{equation}
where the mixing angle $\theta^{ls}_\eta$ ($\theta^{ls}_\sigma$) 
represents the $\eta_{\bar{s}s}$-$\eta_{\bar{l}l}$
($\sigma_{\bar{s}s}$-$\sigma_{\bar{l}l}$) mixture and is obtained
by diagonalizing coupled meson propagators for $\eta_{\bar{l}l}$ and 
$\eta_{\bar{s}s}$ 
($\sigma_{\bar{l}l}$ and $\sigma_{\bar{s}s}$) states~\cite{KHKB}. 
The Fourier transform $\chi_{\xi\xi'} (q^2)$ 
of  mesonic correlation function 
$\zeta_{\xi\xi'} (x) \equiv \langle 0 | {\rm T} \left( J_\xi(x)
J^{\dagger}_{\xi'}(0) \right) | 0 \rangle$ in Minkowski space $x=(t,{\bf
x})$ is described by  
%%%%%%%%
\begin{align}
\chi_{\xi\xi'} (q^2) = \chi_{\xi\xi'} (q^2_0, {\tilde q}^2) 
=  i \int d^4x e^{i q\cdot x} \zeta_{\xi\xi'} (x) 
\end{align}
%%%%%%%%
with (external) 
momentum $q=(q_0,{\bf q})$, where $\tilde{q}=\pm |{\bf q}|$ and 
${\rm T}$ stands for the time-ordered product. Using the random-phase (ring) approximation,
one can obtain the Schwinger-Dyson equation
%%%
\bea
  \chi_{\xi\xi'}
= \Pi_{\xi\xi'} +2 \sum_{\xi''\xi'''} \Pi_{\xi\xi''}G_{\xi''\xi'''}\chi_{\xi'''\xi'} 
\label{SD}
\eea
%%%%
for $\chi_{\xi\xi'}$, where 
$G_{\xi\xi'}$ is an effective four-quark interaction working between 
mesons $\xi$ and $\xi'$. The one-loop polarization function 
$\Pi_{\xi\xi'}$ is  defined by 
%%%%
\bea
  \Pi_{\xi\xi'}(q^2)  \equiv  (-i) \int \frac{d^4 p}{(2\pi)^4} 
  {\rm Tr} \left( \Gamma_\xi iS(p'+q) \Gamma_{\xi'} iS(p') \right)
\label{pi_xi} 
\eea
%%%%%
with internal momentum $p=(p_{0},{\bf p})$, where 
$p'=(p_{0}+iA_4,{\bf p})$ and the trace ${\rm Tr}$ is taken 
in flavor, Dirac and 
color spaces. The quark propagator $S(p)$  
is diagonal in flavor space: $S(p)={\rm diag}(S_u,S_d,S_s)$. The
polarization function $\Pi_{\xi\xi'}(q^2)$ can be classified with quark
and anti-quark flavors $f$ and $f'$ as
%%%%%%%%%%%%%%%%
\begin{eqnarray}
\Pi_{\rm S}^{ff'}
&=&(-2i) \int \frac{d^4 p}{(2\pi)^4} 
{\rm tr}_{\rm c,d} \left(iS_f(p'+q)iS_{f'}(p')\right)
\nonumber\\
&=&4i[I_{1}^{f}+I_{2}^{f'}-\left\{q^2-(M_f + M_{f'})^2\right\}I_{3}^{ff'}] 
\label{Pi_S}
\end{eqnarray}
for the scalar mesons and 
\begin{eqnarray}
 \Pi_{\rm P}^{ff'}
  &=&(-2i) \int \frac{d^4 p}{(2\pi)^4} 
{\rm tr}_{\rm c,d} \left((i\gamma_5)iS_f(p'+q)(i\gamma_5)iS_{f'}(p')\right)
\nonumber\\
  &=&4i[I_{1}^{f}+I_{2}^{f'}-\left\{q^2-(M_f - M_{f'})^2\right\}I_{3}^{ff'}
\label{Pi_P}
\end{eqnarray}
for pseudoscalar mesons, 
where the trace ${\rm tr}_{\rm c,d}$ is taken in color and Dirac
spaces and 
%%%%%%%%%%%%%%%%
\begin{eqnarray}
I_{1}^f&=&\int {d^4p\over{(2\pi )^4}}{\rm tr_c}\Bigl[{1\over{p'^2-M_f^2}}\Bigr],
\label{I1}
\\
I_{2}^f&=&\int {d^4p\over{(2\pi )^4}}{\rm tr_c}\Bigl[{1\over{(p'+q)^2-M_f^2}}\Bigr],
\label{I2}
\\
I_{3}^{ff'}&=&\int {d^4p\over{(2\pi )^4}}{\rm
 tr_c}\Bigl[{1\over{\{p'^2-M^2_f\}((p' + q)^2-M^2_{f'})}}\Bigr] .
\nonumber\\
\label{I3}
\end{eqnarray}
%%%%%%%%%%%%%%
For finite $T$, the corresponding equations are 
obtained by the replacement
%%%%% 
\begin{align}
&p_0 \to i \omega_n = i(2n+1) \pi T, 
\nonumber\\
&\int \frac{d^4p}{(2 \pi)^4} 
\to iT\sum_{n=-\infty}^{\infty} \int \frac{d^3{\bf p}}{(2 \pi)^3}. 
\label{finte_T_mu}
\end{align}
%%%%%

Here we explain the PV regularization for the thermodynamic potential
$\Omega$ of Eq.~\eqref{PNJL-Omega} and the three integrals 
$I_1^{f},I_2^{f},I_3^{ff'}$. For convenience, we divide $\Omega$ into $\Omega =
U_{\rm M} + \mathcal{U} + \sum_{f=u,d,s}\Omega_{\rm F}(M_f)$, and represent $I_{1}^f$ and $I_{2}^f$ by $I(M_f)$ and 
$I_{3}^{ff'}$ by $I_{ff'}(M_f,M_{f'})$. In the PV scheme, the functions
$\Omega_{\rm F}(M_{f})$, $I(M_f)$ and $I_{ff'}(M_f,M_{f'})$ 
are regularized as 
%%%%%%%%%%%%%%%%
\begin{eqnarray}
  \Omega^{{\rm reg}}_{\rm F}(M_f)&=&\sum_{\alpha=0}^2 C_\alpha \Omega_{\rm F}(M_{f;\alpha}),
   \nonumber
   \\
 I^{\rm reg}(M_f)&=&\sum_{\alpha=0}^2 C_\alpha I(M_{f;\alpha}),
  \nonumber
  \\  
 I^{\rm reg}_{ff'}(M_f,M_{f'})&=&\sum_{\alpha=0}^2
  C_\alpha I_{ff'}(M_{f;\alpha},M_{f';\alpha}) ,
\label{PV}
\end{eqnarray}
%%%%%%%%%%%%%%%%
where 
$M_{f;0}=M_f$ and the $M_{f;\alpha}~(\alpha =1, 2)$ mean masses 
of auxiliary particles. The parameters $M_{f;\alpha}$ and $C_\alpha$ 
should satisfy the condition  
$\sum_{\alpha=0}^2C_\alpha=\sum_{\alpha=0}^2 C_\alpha
M_{f;\alpha}^2=0$~to remove the quartic, the quadratic and the logarithmic divergence in
$I_1,I_2,I_3^{ff'}$, and $\Omega_{\rm F}$. Logarithmic
divergence partially remains in $\Omega^{{\rm reg}}_{\rm F}(M_{f})$ 
even after the subtraction of Eq.~\eqref{PV}, 
but the term does not depend on the mean-field variables 
($\sigma_l,\sigma_s,\Phi,\bar{\Phi}$) and is irrelevant 
to the determination of mean-field variables for any $T$. 
Therefore we can simply drop the term. 
We assume $(C_0,C_1,C_2)=(1,1,-2)$ and 
$(M_{f;1}^2,M_{f;2}^2)=(M^2_{f}+2\Lambda^2,M^2_{f}+\Lambda^2)$,
following Ref.~\cite{Itzykson;Zuber}. We keep the parameter $\Lambda$ finite 
even after the subtraction \eqref{PV}, since the present model is 
non-renormalizable.

%%%%%
\subsubsection{$\pi,a_0,K,\kappa$ mesons}
%%%%%
For $\xi=\pi, a_0, K$ and $\kappa$, the effective four-quark interactions 
$G_{\xi\xi'}$ and the polarization functions $\Pi_{\xi\xi'}$ are diagonal, 
i.e., $G_{\xi\xi'}=G_{\xi}\delta_{\xi\xi'}$, $\Pi_{\xi\xi'}=\Pi_{\xi}\delta_{\xi\xi'}$, because of isospin symmetry in the light-quark sector and 
the random-phase approximation. One then can easily 
get the solution to the Schwinger-Dyson equation \eqref{SD} as  
%%%%%
\bea
\chi_{\xi\xi} &=& 
\frac{\Pi_{\xi}}{1 - 2G_{\xi}\Pi_{\xi}}
\label{chi_pia0}  
\eea
%%%%%
for $\xi=\pi, a_0, K$ and $\kappa$, where 
the effective couplings $G_{\xi}$ are 
defined by 
%%%%%%%%%%%%%%%
\begin{eqnarray}
G_{a_0} &=& G_{\rm S}(T) + \frac{1}{2}G_{\rm D}(T) \sigma_s, 
\label{g_a0}
\\
G_{\pi} &=& G_{\rm S}(T) - \frac{1}{2}G_{\rm D}(T) \sigma_s,
\label{g_pi}
\\
G_{\kappa} &=& G_{\rm S}(T) + \frac{1}{2}G_{\rm D}(T) \sigma_l, 
\label{g_kappa}
\\
G_{K} &=& G_{\rm S}(T) - \frac{1}{2}G_{\rm D}(T) \sigma_l
\label{g_K}
\end{eqnarray}
%%%%%%%%%%%%%%
and the one-loop polarization functions $\Pi_{\xi}$ are written by
%%%%%%%%%%%%%%%
\begin{eqnarray}
 \Pi_{a_0}&=&\Pi_{\rm S}^{ll},~~
  \Pi_{\pi}=\Pi_{\rm P}^{ll},~~
\Pi_{\kappa}=\Pi_{\rm S}^{sl},~~
 \Pi_{K}=\Pi_{\rm P}^{sl}.\nonumber\\
\end{eqnarray}
%%%%%%%%%%%%%%%

The meson pole mass $M_{\xi}^{\rm pole}$ is a pole of 
$\chi_{\xi\xi}(q_0^2,{\tilde q}^2)$ in the complex $q_0$ plane. 
Taking the rest frame $q=(q_0,{\bf 0})$ for convenience, 
one can get the equation  
%%%%%
\begin{align}
\big[1 - 2G_{\xi}\Pi_{\xi}(q_0^2,0)\big]\big|_{q_0=M_{\xi}^{\rm pole}-i\Gamma_\xi/2}=0    
\label{mmf}
\end{align}
%%%%%
for $M_{\xi}^{\rm pole}$, where $\Gamma_\xi$ is the decay width to $q\bar{q}$ 
continuum. The $M_{\xi}^{\rm pole}$ and $\Gamma_\xi$ are obtained numerically by searching for 
the $q_0$ satisfying Eq. \eqref{mmf}. Here we take the approximation
$\Gamma_{\xi}/M_{\xi}^{\rm pole}\ll 1$, following Ref.~\cite{Costa:2005}.

%%%%%
\subsubsection{$\eta,\eta',\sigma,f_0$ mesons}
%%%%%
The pole masses of $\eta$ and $\eta'$ ($\sigma$ and $f_0$) mesons are obtained 
by solving the coupled-channel equations (\ref{SD}) for $\eta_{\bar{l}l}$
and $\eta_{\bar{s}s}$ ($\sigma_{\bar{l}l}$ and $\sigma_{\bar{s}s}$). 
It is convenient to express 
the correlation functions $\chi_{\xi\xi'}$ with the matrix 
\begin{eqnarray}
{\boldsymbol \chi}_{\xi}
=
\left(
\begin{array}{cc}
\chi_{\xi_{\bar{s}s}\xi_{\bar{s}s}}& \chi_{\xi_{\bar{s}s}\xi_{\bar{l}l}}\\
\chi_{\xi_{\bar{l}l}\xi_{\bar{s}s}}& \chi_{\xi_{\bar{l}l}\xi_{\bar{l}l}}
\end{array}
 \right)~~(\xi =\eta,\sigma).  
\label{correlation-function-matrix}
\end{eqnarray}
The Schwinger-Dyson equation for $\boldsymbol{\chi}_{\xi}$ is obtained from 
Eq. \eqref{SD} as
\begin{equation}
\boldsymbol{\chi}_\xi =  \boldsymbol{\Pi}_\xi +
 2\boldsymbol{\Pi}_\xi\boldsymbol{G}_\xi\boldsymbol{\chi}_\xi
 \label{SD_mix}
\end{equation}
with 
\begin{eqnarray}
{\boldsymbol G}_{\xi}
&=&
\left(
\begin{array}{cc}
G_{\xi_{\bar{s}s}\xi_{\bar{s}s}}& G_{\xi_{\bar{s}s}\xi_{\bar{l}l}}\\
G_{\xi_{\bar{l}l}\xi_{\bar{s}s}}& G_{\xi_{\bar{l}l}\xi_{\bar{l}l}}
\end{array}
\right),~~
{\boldsymbol \Pi}_{\xi}
=
\left(
\begin{array}{cc}
\Pi_{\xi_{\bar{s}s}}& 0\\
0& \Pi_{\xi_{\bar{l}l}}
\end{array}
\right).\nonumber\\
\label{gpi_eta}
\end{eqnarray}
The solution to Eq. \eqref{SD_mix} is   
\begin{eqnarray}
 \chi_{\xi_{\bar{s}s}\xi_{\bar{s}s}}
  &=&\frac{(1-2G_{\xi_{\bar{l}l}\xi_{\bar{l}l}}\Pi_{\xi_{\bar{l}l}})\Pi_{\xi_{\bar{s}s}}}
{\det{\left[{\boldsymbol I} - 2{\boldsymbol \Pi}_{\xi}{\boldsymbol
       G}_{\xi}\right]}},
\label{chi_ss_p}
\\
 \chi_{\xi_{\bar{l}l}\xi_{\bar{l}l}}
  &=&\frac{(1-2G_{\xi_{\bar{s}s}\xi_{\bar{s}s}}\Pi_{\xi_{\bar{s}s}})\Pi_{\xi_{\bar{l}l}}}
{\det{\left[{\boldsymbol I} - 2{\boldsymbol \Pi}_{\xi}{\boldsymbol
       G}_{\xi}\right]}},
\label{chi_ll_p}
\\
 \chi_{\xi_{\bar{s}s}\xi_{\bar{l}l}}
  &=& \chi_{\xi_{\bar{l}l}\xi_{\bar{s}s}}
  =\frac{2G_{\xi_{\bar{l}l}\xi_{\bar{s}s}}\Pi_{\xi_{\bar{s}s}}\Pi_{\xi_{\bar{l}l}}}
{\det{\left[{\boldsymbol I} - 2{\boldsymbol \Pi}_{\xi}{\boldsymbol
       G}_{\xi}\right]}},
\label{chi_sl_p}
\end{eqnarray}
where ${\boldsymbol I}$ is the unit matrix and the determinant $\det$ 
is taken in the $\xi_{\bar{l}l}$ and $\xi_{\bar{s}s}$ channels. 
The matrix elements of $\boldsymbol{G}_\eta$ and $\boldsymbol{G}_{\sigma}$ are
explicitly obtained by 
\begin{eqnarray}
G_{\eta_{\bar{s}s}\eta_{\bar{s}s}}   &=& G_{\rm S}(T),~~
G_{\eta_{\bar{l}l}\eta_{\bar{l}l}} = G_{\rm S}(T) + \frac{1}{2}G_{\rm D}(T)\sigma_{s}, \nonumber\\
G_{\eta_{\bar{s}s}\eta_{\bar{l}l}}  &=&
 G_{\eta_{\bar{l}l}\eta_{\bar{s}s}} = \frac{\sqrt{2}}{2}G_{\rm
 D}(T)\sigma_l,\\
\label{gelement_eta}
G_{\sigma_{\bar{s}s}\sigma_{\bar{s}s}}   &=& G_{\rm S}(T),~~
G_{\sigma_{\bar{l}l}\sigma_{\bar{l}l}} = G_{\rm S}(T) - \frac{1}{2}G_{\rm D}(T)\sigma_{s}, \nonumber\\
G_{\sigma_{\bar{s}s}\sigma_{\bar{l}l}}  &=&
 G_{\sigma_{\bar{l}l}\sigma_{\bar{s}s}} = -\frac{\sqrt{2}}{2}G_{\rm
 D}(T)\sigma_l,  
\label{gelement_sigma}
\end{eqnarray}
and those of $\boldsymbol{\Pi}_{\eta},\boldsymbol{\Pi}_{\sigma}$ are by  
\begin{eqnarray}
 \Pi_{\sigma_{\bar{l}l}}&=&\Pi_{\rm S}^{ll},~~
 \Pi_{\sigma_{\bar{s}s}}=\Pi_{\rm S}^{ss},\\
 \Pi_{\eta_{\bar{l}l}}&=&\Pi_{\rm P}^{ll},~~  
 \Pi_{\eta_{\bar{s}s}}=\Pi_{\rm P}^{ss}.
\end{eqnarray}
%%%%%%%%%%%%%%
The masses of $\eta$ and $\eta'$ ($\sigma$ and $f_0$) are
determined as poles of $\boldsymbol{\chi}_{\eta}$ ($\boldsymbol{\chi}_{\sigma}$), that is, as zero points of the determinant in Eqs. 
(\ref{chi_ss_p})-(\ref{chi_sl_p}):
%%%%%
\begin{align}
 \det{\big[{\boldsymbol I}
 - 2{\boldsymbol \Pi}_\xi(q_0^2,0)~{\boldsymbol G}_\xi\big]\big|_{q_0=M_{\xi}^{\rm pole}-i\Gamma_\xi/2}}=0 .   
\label{mmf2}
\end{align}
%%%%%
Two poles  are found in the complex $q_0$ plane. 
The lighter and heavier pole masses correspond to $\eta$ 
and $\eta'$ ($\sigma$ and $f_0$) meson masses, respectively.

\subsection{Meson screening masses}
\label{Meson screening masses}

We first show the reason why the derivation of
$M_{\xi}^{\rm scr}(T)$ 
was difficult in NJL-type effective models before the work 
of Ref.~\cite{Ishii:2013kaa}, and next recapitulate the method of 
Ref.~\cite{Ishii:2013kaa} and extend it from single-channel systems to 
multi-channel systems.
The $M_{\xi}^{\rm scr}$
is defined with 
the spatial correlator $\zeta_{\xi\xi}(0,{\bf x})$ 
in the long-distance limit ($r=|\bf x|\to\infty$):
\begin{equation}
M_{\xi}^{\rm scr}
=-\lim_{r\rightarrow \infty}\frac{d \ln{\zeta_{\xi\xi}(0,{\bf x})}}{dr}, 
\label{scr-mass}
\end{equation}
where 
\begin{equation}
\zeta_{\xi\xi}(0,{\bf
 x})=\frac{1}{4\pi^{2}ir}\int^{\infty}_{-\infty}d\tilde{q}\hspace{1ex}\tilde{q}\chi_{\xi\xi}(0,\tilde{q}^2)e^{i\tilde{q}r} .
\label{chi_r}
\end{equation}
%%%%%%%%%%%%%%
Equation \eqref{chi_r} has two problems, 
when the $\tilde{q}$ integration is performed.  
The first problem stems from the regularization taken. 
As already mentioned in Sec. \ref{Model setting}, 
the 3d momentum cutoff is commonly used, but it breaks Lorentz
invariance even in $T=0$. This induces an unphysical oscillation in 
$\zeta_{\xi\xi}(0,{\bf x})$ at large $r$ \cite{Florkowski}. 
We can easily solve this problem by using the PV regularization. 
This is the reason why we take the PV regularization in this paper. 
As easily found from Eq. \eqref{chi_r}, direct numerical calculations 
of the $\tilde{q}$ integration are quite difficult at large $r$ 
because of highly oscillation of the integrand. 
This is the second problem. 
In order to solve this problem, one can consider 
analytic continuation of $\chi_{\xi\xi}(0,{\tilde{q}^2})$ 
to the  complex $\tilde{q}$ plane. 
In general, the integration can be made easily 
with the Cauchy's integral theorem. However, 
the complex function $\chi_{\xi\xi} (0,\tilde{q}^{2})$ has logarithmic cuts 
in the vicinity of the real $\tilde{q}$ axis \cite{Florkowski}.  This
demands quite 
time-consuming numerical calculations to evaluate the contribution 
of logarithmic cuts \cite{Florkowski}. 
In our previous works~\cite{Ishii:2013kaa,Ishii:2015ira}, we showed 
that these logarithmic cuts are not physical and avoidable by taking
the Matsubara summation over $n$ after the ${\bf p}$ integration 
in Eq. \eqref{finte_T_mu}. 
Consequently, we obtain the regularized function $I_{3,{\rm reg}}^{ff'}$ as 
an infinite series of analytic functions:  
%%%%%%%%%%%%%%
\begin{eqnarray}
&&I_{3,{\rm reg}}^{ff'}(0,\tilde{q}^{2})=iT\sum_{j=1}^{N_c}\sum_{n=-\infty}^\infty\sum_{\alpha=0}^2C_{\alpha}
\nonumber\\
&\times & \int {d^3{\bf p}\over{(2\pi )^3}}
\Bigl[{1\over{{\bf p}^2+\mathcal{M}_f^2}}{1\over{({\bf p + q})^2+\mathcal{M}_{f'}^2}}\Bigr]\nonumber\\
&=&{iT\over{2\pi^2}}\sum_{j,n,\alpha}C_{\alpha}\int_0^1dx\int_0^\infty
 dk\nonumber\\
 &\times &{k^2\over{[k^2+(x-x^2)\tilde{q}^2+(1-x)\mathcal{M}_f^2 + x\mathcal{M}_{f'}^2]^2}}
\nonumber\\
 &=&{T\over{8\pi \tilde{q}}}\sum_{j,n,\alpha}C_{\alpha}
  {\rm Log}\left(\frac{\mathcal{M}_f + \mathcal{M}_{f'} + i\tilde{q}}
            {\mathcal{M}_f + \mathcal{M}_{f'} - i\tilde{q}}\right)
\label{I_3_final}
\end{eqnarray}
%%%%%%%%%%%%%%
with 
%%%%%%%%%%%%%%%%
\begin{equation}
\mathcal{M}_f(T)=\sqrt{M_{f,\alpha}^2 + \{(2n+1)\pi T+A_{4}^{jj}\}^2 }, 
\label{KK_mode}
\end{equation}
where ``${\rm Log}$'' denotes the
principle value of the logarithm. The function $iI_{3,{\rm reg}}^{ff'}$ is 
real for real $\tilde{q}$, when $q_0=0$. This means 
that mesons do not decay into a quark and an antiquark. 
The function $I_{3,{\rm reg}}^{ff'}$ is obtained as 
an infinite series, but we numerically confirmed that 
the sequence of partial sums converges rapidly. 
In the last form of Eq. (\ref{I_3_final}), 
each term has two physical cuts on the imaginary axis; 
one is an upward vertical line 
starting from the branch point $\tilde{q} = i\left(\mathcal{M}_{f} + \mathcal{M}_{f'}\right)$ and the other is 
a downward vertical line from the branch point $\tilde{q} = -
i\left(\mathcal{M}_{f} + \mathcal{M}_{f'}\right)$. In the upper half-plane
where the contour integration is taken, the lowest branch point is $\tilde{q}=i\left(\mathcal{M}_{f} +
\mathcal{M}_{f'}\right)_{j=1,n=0,\alpha=0}$.

The screening mass $M_{\xi}^{{\rm scr}}$ is determined as a pole of
$\chi_{\xi\xi}(0,\tilde{q}^2)$ on the imaginary $\tilde{q}$ axis.  
The pole should be located below the lowest branch point: 
\bea
M_{\xi}^{\rm scr}<
M_{\rm th} \equiv 
\left(\mathcal{M}_{f} +
\mathcal{M}_{f'}\right)_{j=1,n=0,\alpha=0}, 
\eea 
where $M_{\rm th}$ can be regarded as  
``threshold mass'' in the sense that meson is in $q\bar{q}$ continuum states 
when $M_{\xi}^{\rm scr}>M_{\rm th}$. 
For $\xi=\pi,a_0,K,\kappa$ channels, 
we can obtain the $M_{\xi}^{{\rm scr}}$ by
solving the equation
%%%%%%%%%%%%%
\begin{align}
\big[1 - 2G_{\xi} \Pi_{\xi}(0,\tilde{q}^2)\big]\big|_{\tilde{q}=iM_{\xi}^{{\rm scr}}}=0  , 
\label{meson_screening}
\end{align}
%%%%%%%%%%%
when $M_{\xi}^{\rm scr}<M_{\rm th}$. 
As $T$ increases, $M_{\xi}^{\rm scr}$ (pole) approaches 
$M_{\rm th}$ (the lowest branch point) 
from below~\cite{Ishii:2013kaa,Ishii:2015ira}. 
Meanwhile, $M_{\rm th}$ itself tends to $2\pi T$ in the high-$T$ limit, 
since $A_4^{jj}$ does to 0 in Eq. \eqref{KK_mode}. 
Therefore, $M_{\xi}^{\rm scr}$ approaches $2\pi T$ 
with respect increasing $T$.

Now we consider the channel mixing. 
The formalism on meson screening masses is the same as that on meson 
pole masses. Only the difference is that the external momentum is 
set to $q=(0,{\bf q})$. The coupled equations for 
the $M_{\xi}^{\rm scr}$ are 
%%%%%
\begin{align}
 \det{\big[{\boldsymbol I}
 - 2{\boldsymbol \Pi}_\xi(0,\tilde{q}^2)~{\boldsymbol G}_\xi\big]
\big|_{\tilde{q}=iM_{\xi}^{{\rm scr}}}}=0  , 
\label{mmf2-scr}
\end{align}
%%%%%
where $\tilde{q} =\pm |{\bf q}|$. Here note that 
$M_{\rm th} = 2(\mathcal{M}_{l})_{j=1,n=0,\alpha=0}$ for
$\eta,\sigma$ mesons. 
For $\eta',f_0$ mesons, we consider 
$M_{\rm th} = 2(\mathcal{M}_{s})_{j=1,n=0,\alpha=0}$ 
as the threshold mass. Strictly speaking, $\eta'$
($f_0$) can decay into a light-quark pair by the 
channel mixing. However, such a contribution is unphysical for low temperature
because of the color confinement and small for high temperature because of
small channel mixing.

\subsection{Model tuning for LQCD-data analyses}
\label{Model tuning for LQCD-data analyses}

We use LQCD data of Ref.~\cite{Cheng:2010fe} for the $M_{\xi}^{\rm scr}(T)$ 
and of Ref.~\cite{Cheng:2008} for $\Delta_{l,s}(T)$, 
since the same lattice setting is taken 
in the two simulations. 
In Refs.~\cite{Cheng:2008,Cheng:2010fe}, the quark-mass ratio is $m_l/m_s=1/10$, and 
the $\pi$-meson mass at $T=0$ is $M_{\pi}^{\rm pole}(0)=176$ MeV that is 
slightly heavier than the experimental value $138$ MeV. 
In model calculations, we then change 
quark masses from $(m_l,m_s)=(8~{\rm MeV},191~{\rm MeV})$ to 
$(m_l,m_s)=(13~{\rm MeV},130~{\rm MeV})$  
to become consistent with the lattice setting. 
This parameter set is tabulated as set (B) in 
Table \ref{Other model parameters}.

In LQCD simulations of 
Refs.~\cite{Cheng:2008,Cheng:2010fe}, $T_c^\chi$ is measured to be
$196$ MeV, but the value established in state-of-art LQCD simulations of 
Refs.~
\cite{Borsanyi:2010bp,Bazavov:2011nk} is $T_c^\chi=154\pm 9$
MeV. Therefore, we rescale the values of $T$ and $M_{\xi}^{\rm scr}$ in
Refs.~\cite{Cheng:2008,Cheng:2010fe} to reproduce $T_c^{\chi}=154\pm
9$ MeV.

In LQCD simulations of Ref.~\cite{Cheng:2010fe} for 
pseudoscalar mesons $(\eta,\eta')$  and scalar ones $(\sigma,f_0)$, 
the quark-line disconnected diagrams are neglected and thereby 
the $\eta_{\bar{s}s}$ 
($\sigma_{\bar{s}s}$)  channel is decoupled with 
the $\eta_{\bar{l}l}$ ($\sigma_{\bar{l}l}$)  channel. 
Eventually, LQCD data are available only for 
$\eta_{\bar{s}s}$- and $\sigma_{\bar{s}s}$-meson screening masses. 
We then switch off the channel mixing in model calculations 
by setting 
$G_{\xi_{\bar{s}s}\xi_{\bar{l}l}}=G_{\xi_{\bar{l}l}\xi_{\bar{s}s}}=0$  
for $\xi=\eta,\sigma$, when we analyze the LQCD data 
on $\eta_{\bar{s}s}$ and $\sigma_{\bar{s}s}$ mesons.

Particularly for $\eta$- and $\eta'$-meson masses at $T=0$, 
it is shown in Ref.~\cite{Christ:2010} that the disconnected 
diagrams are necessary to reproduce the experimental values, 
although they are neglected in finite-$T$ LQCD simulations 
of Ref.~\cite{Cheng:2010fe} for $M_{\xi}^{\rm scr}(T)$. 
The disconnected diagrams contribute to both diagonal and 
off-diagonal elements of 
the correlation-function matrix ${\boldsymbol \chi}_{\xi}$ 
in Eq. \eqref{correlation-function-matrix}, 
whereas the connected diagrams do to only the diagonal elements. 
The channel mixing induced by the off-diagram elements
 is thus one of effects induced 
by the disconnected diagrams. 
We can then divide the disconnected-diagrams effects into 
the channel-mixing effect and 
the remaining disconnected-diagram effects acting on the diagonal elements of 
${\boldsymbol \chi}_{\xi}$. 
Model calculations with the parameter set (A) include 
the channel-mixing effect explicitly and 
the remaining disconnected-diagram effects implicitly, since 
the set (A) is so determined as to reproduce experimental data on 
meson pole masses at $T=0$, particularly on  $M_{\eta'}^{\rm pole}(0)$. 
Hence, we can consider that model calculations with the parameter 
set (B) also include the channel-mixing effect explicitly and 
the remaining disconnected-diagram effects implicitly, 
whereas LQCD calculations do not have any disconnected-diagram effects. 
We switch off the channel mixing in model calculations to 
evaluate $\eta_{\bar{s}s}$- and $\sigma_{\bar{s}s}$-meson screening masses, 
but we should note that the remaining disconnected-diagram effects are 
included in model calculations implicitly.

%%%%%%%%%%%%%%%%%%%%%%%%%%%%%%%%%%%%%%%%%%%%%%%%%%%%%%%%%%%%%%%%%%%%%%%%%%%
%%%%% Numerical results
%%%%%%%%%%%%%%%%%%%%%%%%%%%%%%%%%%%%%%%%%%%%%%%%%%%%%%%%%%%%%%%%%%%%%%%%%%% 
\section{Numerical Results}
\label{Numerical Results}

\subsection{Parameter fitting}
\label{Parameter fitting}

As shown in Eqs. \eqref{T-dependent-K} and \eqref{T-dependent-G}, 
the present model has adjustable parameters $(T_1,b_1)$ in the KMT 
coupling strength $G_{\rm D}(T)$ and $(T_2,b_2)$ 
in the four-quark coupling strength $G_{\rm S}(T)$. 
The parameters $(T_1,b_1)$ are determined from 
LQCD data associated with 
the $U(1)_{\rm A}$-symmetry restoration, i.e., 
$\Delta M_{a_0,\pi}^{\rm scr}= M_{a_0}^{\rm scr}- M_{\pi}^{\rm scr}$ 
in $T > 1.1T_c^\chi=170$~MeV 
where only the $U(1)_{\rm A}$-symmetry breaking survives~\cite{Bhattacharya:2014ara,Ishii:2015ira}. 
Similarly, the parameters $(T_2,b_2)$ are determined from  LQCD data 
associated with the chiral-symmetry restoration, i.e., 
the pseudocritical temperature $T_c^{\chi}=154\pm 9$ MeV~\cite{Borsanyi:2010bp,Bazavov:2011nk} and the renormalized chiral 
condensate $\Delta_{l,s}(T)$ of Eq. \eqref{eq:renormalized chiral condensate}.

\begin{figure}[t]%[H]
\begin{center}
   \includegraphics[width=0.4\textwidth]{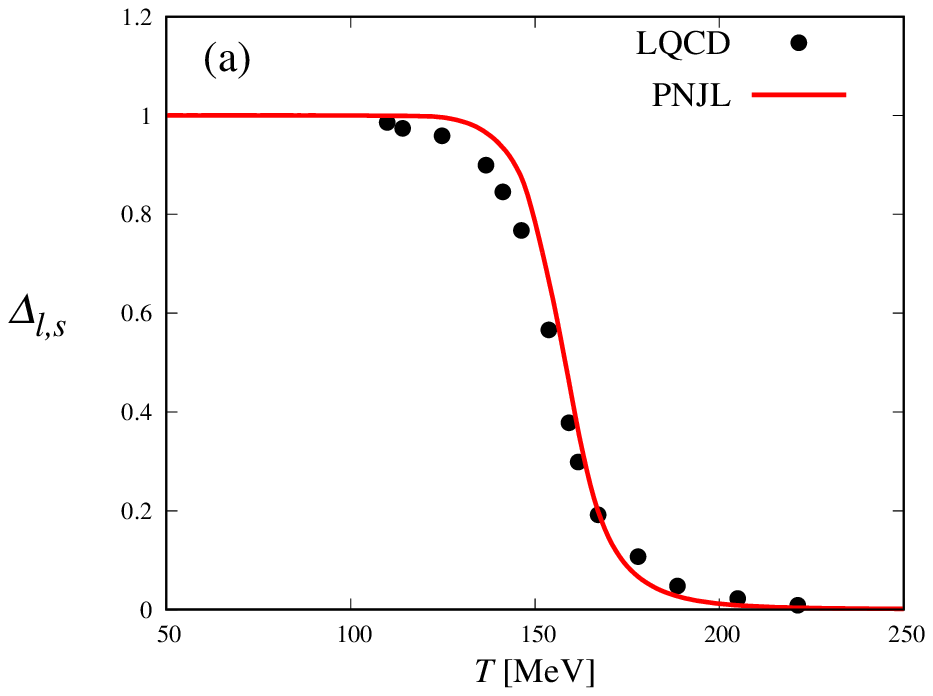}
   \includegraphics[width=0.43\textwidth]{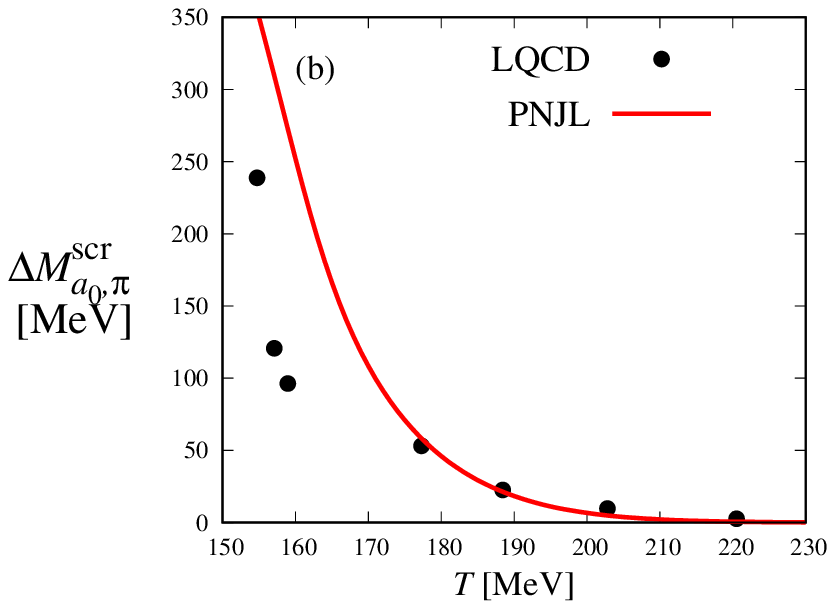}
\end{center}
\caption{$T$ dependence 
of (a) $\Delta_{l,s}$ and (b) $\Delta M_{a_0,\pi}^{\rm scr}$.
Model results are shown by solid lines, while LQCD data are denoted 
by closed circles. 
The parameter set (B) is taken in model calculations. 
LQCD data are taken from Refs.~\cite{Cheng:2008,Cheng:2010fe}. 
}
\label{fitting}
\end{figure}

Figure \ref{fitting} shows the results of the present parameter fitting 
for (a) $\Delta_{l,s}(T)$ and (b) $\Delta M^{\rm scr}_{a_0,\pi}(T)$. 
Note that the parameter set (B)  is 
taken in model calculations. Good agreement 
is seen between model results (solid lines) and LQCD data (closed circles), 
when $(T_1,b_1)=(121, 43.5)$ and $(T_2,b_2)=(131, 83.3)$ 
in units of MeV. These values are tabulated 
in Table \ref{Model parameters in coupling strengths}.

 %%%%%%%%%%%%%%%% Fig %%%%%%%%%%%%%%%%%%%%%
%%%%%%%%%%%%%

\subsection{Meson screening masses}
\label{results:Meson screening masses}

We consider meson screening masses for pseudoscalar and scalar mesons and 
analyze LQCD data of Ref.~\cite{Cheng:2010fe}, 
using the present model with the parameter set (B). 
In the model calculations the channel mixing is switched off, 
since the disconnected diagrams are neglected in 
LQCD data of Ref.~\cite{Cheng:2010fe}.

%%%%%%%%%%%%%%%% Fig %%%%%%%%%%%%%%%%%%%%%
\begin{figure}[t]%[H]
\begin{center}
   \includegraphics[width=0.45\textwidth]{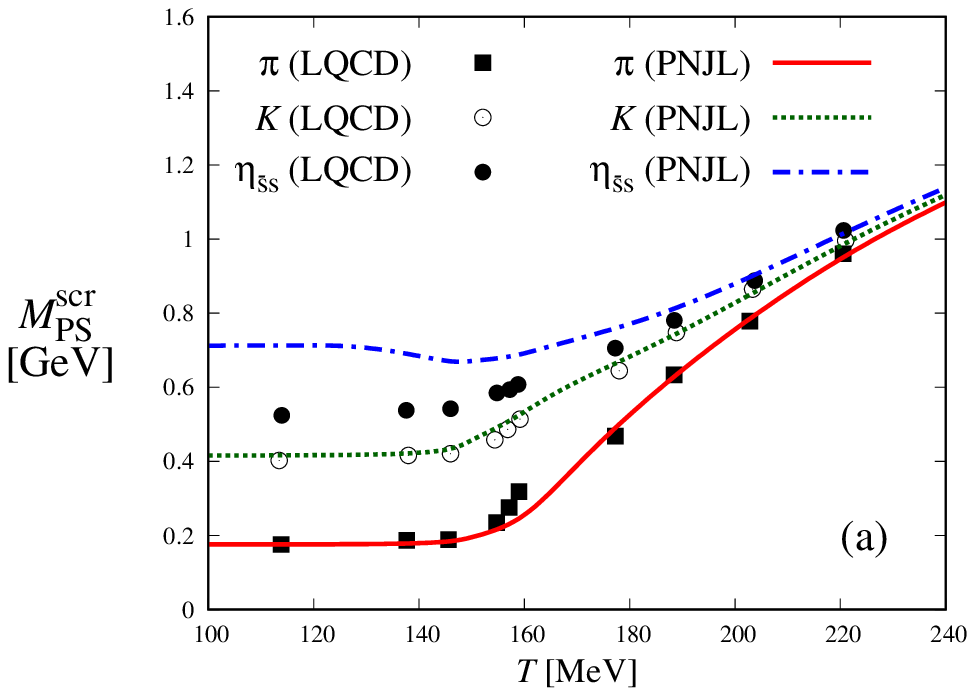}
   \includegraphics[width=0.45\textwidth]{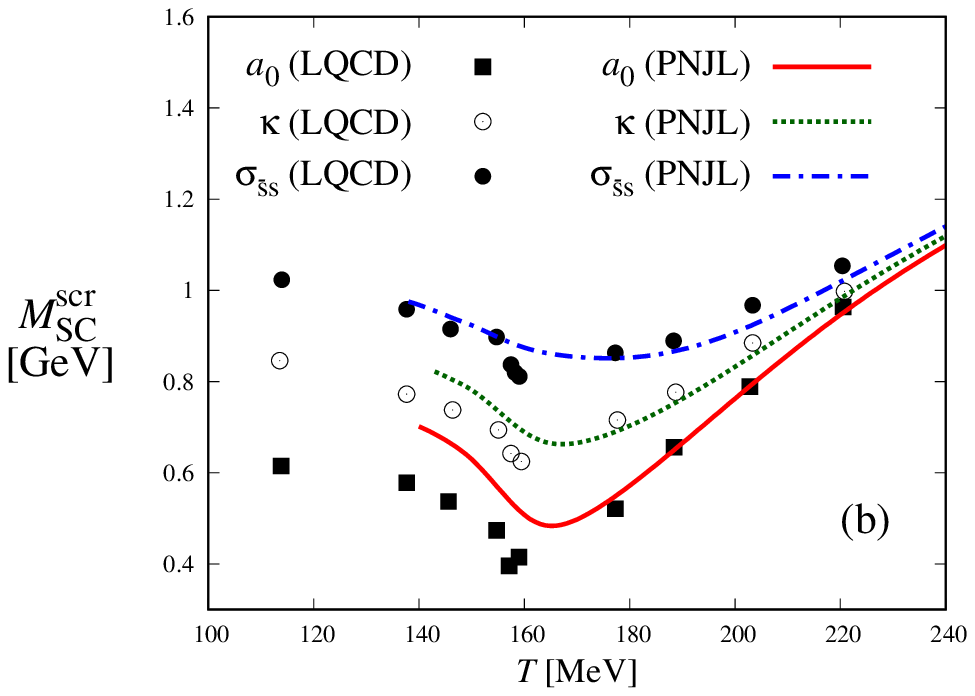}
\end{center}
 \caption{$T$ dependence of meson screening masses 
for (a) pseudoscalar mesons $\pi,K,\eta_{\bar{s}s}$ and (b) scalar mesons 
$a_0,\kappa,\sigma_{\bar{s}s}$. 
Model results are denoted by lines and LQCD data are by symbols. 
The parameter set (B) is taken in model calculations. 
LQCD data are taken from from Ref.~\cite{Cheng:2010fe}. }
\label{scr_mass}
\end{figure}
%%%%%%%%%%%%%

Figure \ref{scr_mass} shows $T$ dependence of 
the $M_{\xi}^{\rm scr}(T)$  for (a) pseudoscalar mesons 
$\xi=\pi,K,\eta_{\bar{s}s}$ and (b) scalar mesons 
$\xi=a_0,\kappa,\sigma_{\bar{s}s}$. 
The lines stand for model results, and the symbols correspond to 
LQCD data of Ref.~\cite{Cheng:2010fe}. 
As mentioned in Sec. \ref{Meson screening masses}, 
in model calculations the $M_{\xi}^{\rm scr}(T)$ are derivable 
when $M_{\xi}^{\rm scr}(T)<M_{\rm th}$. For the $a_0$-meson case, for example, 
the condition is satisfied for $T>139$~MeV. 
The solid line representing $M_{a_0}^{\rm scr}(T)$ 
is then drawn in $T>139$~MeV. 
The same procedure is taken for the other lines. 
In both LQCD data and model results, all the meson masses 
tend to $2\pi T$ with respect to increasing $T$; 
see Sec. \ref{Meson screening masses} for the proof. 
Owing to this property, in $T > 1.04T_c^{\chi}=160$ MeV, 
model results well reproduce LQCD data  for all the mesons. 
In $T < 1.04T_c^{\chi}=160$ MeV, the 
agreement between model results and LQCD data is good for 
pseudoscalar $\pi,K$ mesons and pretty good for scalar 
$a_0,\kappa,\sigma_{\bar{s}s}$ mesons. For pseudoscalar $\eta_{\bar{s}s}$ meson, the model result overestimates 
LQCD data by about $10\% \sim 30\%$ 
in $T < 1.04T_c^{\chi}=160$ MeV, but the deviation 
becomes small rapidly as $T$ increases from $160$ MeV. 
The deviation in $T < 1.04T_c^{\chi}=160$ MeV 
may come from 
the remaining disconnected-diagram effects acting on the diagonal elements of 
${\boldsymbol \chi}_{\xi}$. 
This implies that the channel-mixing effect is also important 
for $\eta_{\bar{s}s}$ meson in $T <  1.04T_c^{\chi}=160$ MeV. 
This statement is confirmed with model calculations 
in Sec. \ref{Discussions}. In addition, this statement is consistent with the statement of Ref. \cite{Bazavov:2012} that 
the disconnected diagrams may be suppressed at least 
for $T \gg T_c^{\chi}$ by the Debye screening 
and the weakly interacting nature of the deconfined phase.

\begin{figure}[t]%[H]
\begin{center}
   \includegraphics[width=0.45\textwidth]{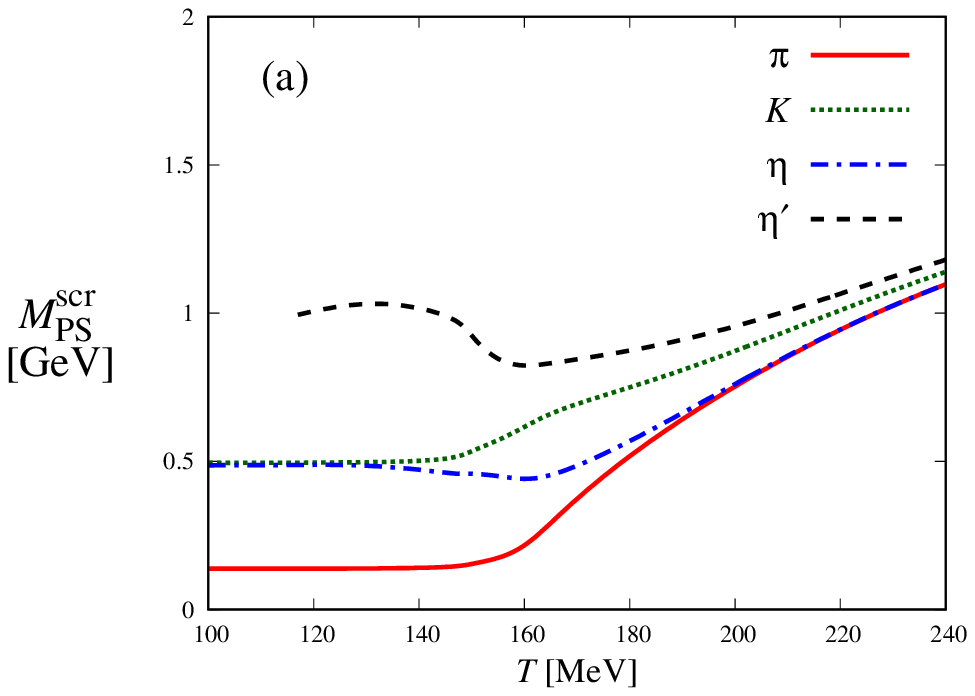}
   \includegraphics[width=0.45\textwidth]{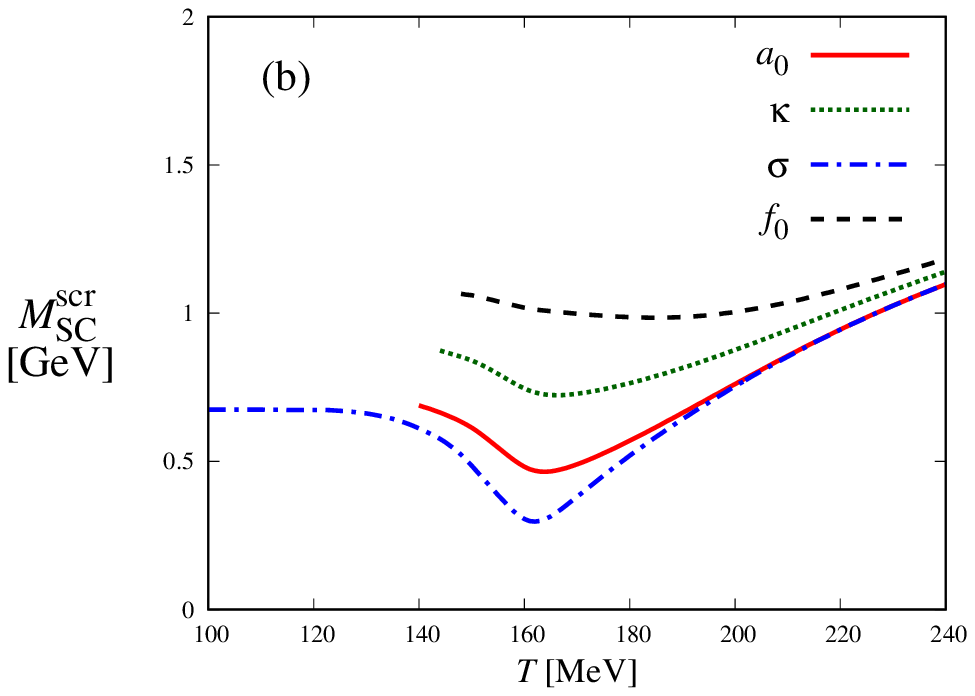}
\end{center}
 \caption{
$T$ dependence of 
meson screening masses 
for (a) pseudoscalar mesons $\pi,K,\eta,\eta'$ and (b) scalar mesons 
$a_0,\kappa,\sigma,f_0$ calculated with the realistic parameter set (A). 
Model results are denoted by lines. 
In model calculations, the channel mixing 
is taken into account. 
}
\label{scr_mass-realistic}
\end{figure}

For later discussion, we evaluate the $M_{\xi}^{\rm scr}(T)$ 
also in the realistic case, taking the parameter set (A) 
and taking account of the channel mixing in model calculations. 
Figure \ref{scr_mass-realistic} shows the results for 
(a) pseudoscalar mesons $\pi,K,\eta,\eta'$ and (b) scalar mesons 
$a_0,\kappa,\sigma,f_0$. 
As mentioned in Fig. \ref{scr_mass}, all the meson screening masses 
tend to $2\pi T$. 
This property is independent of quark masses. At high $T$, 
the $M_{\xi}^{\rm scr}(T)$ calculated with the realistic parameter set (A) 
are close to those with the set (B). 
The difference between the former and the 
latter appear only in $T < T_c^\chi=154\pm9$~MeV.

%%%%%%%%%%%%%%%% Fig %%%%%%%%%%%%%%%%%%%%%
%%%%%%%%%%%%%

\subsection{Meson pole masses}
\label{results: Meson pole masses}

\begin{figure}[t]%[H]
\begin{center}
   \includegraphics[width=0.45\textwidth]{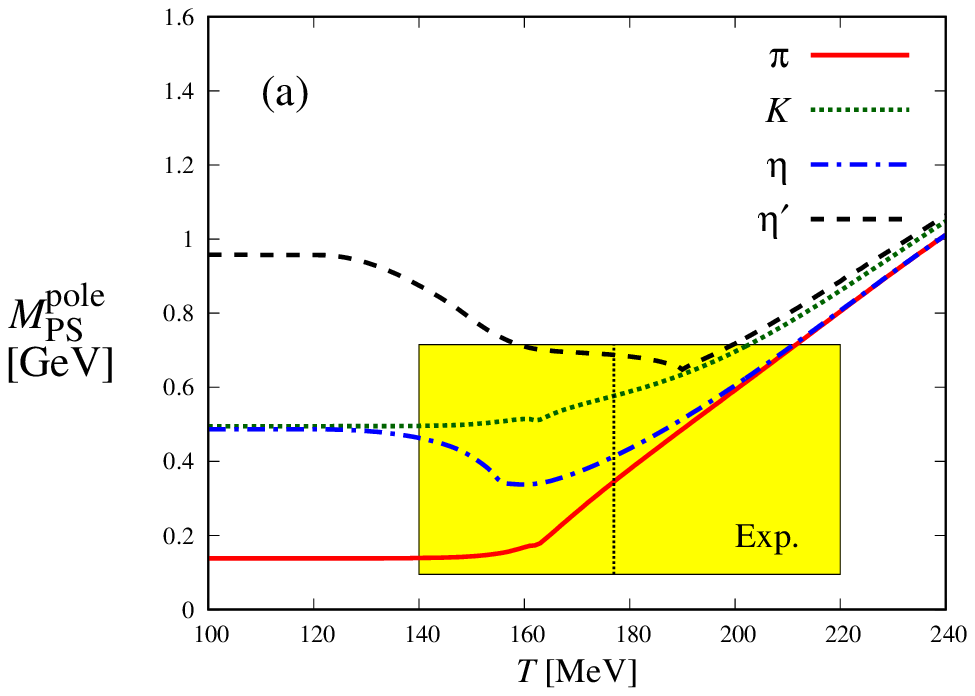}
   \includegraphics[width=0.45\textwidth]{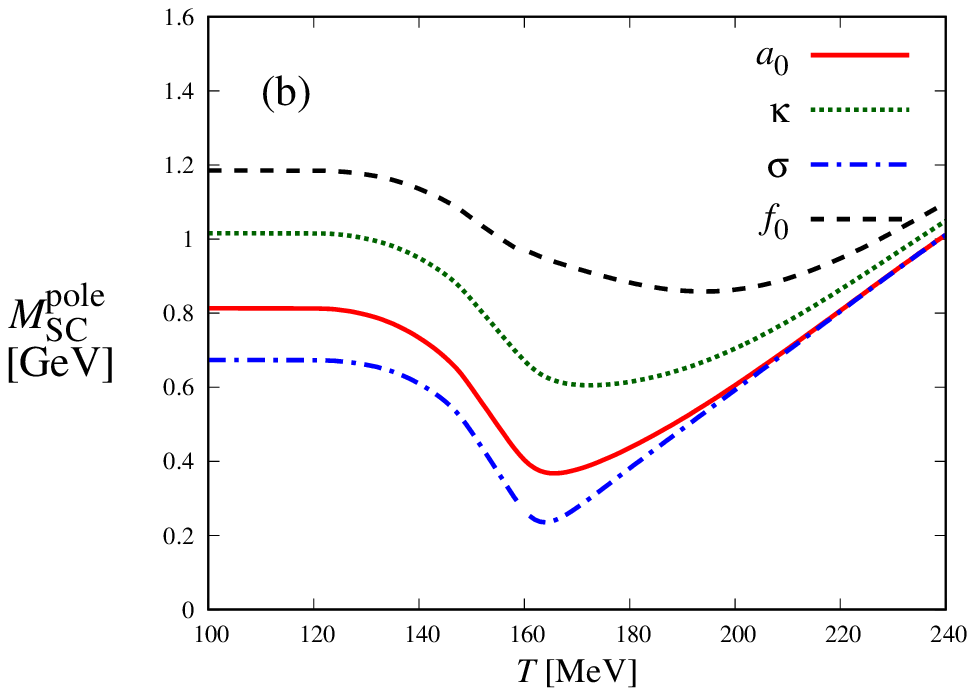}
\end{center}
 \caption{ 
Model prediction on $T$ dependence of meson pole masses 
for (a) pseudoscalar mesons 
$\pi,K,\eta,\eta'$ and (b) scalar mesons 
$a_0,\kappa,\sigma,f_0$. 
In model calculations, the parameter set (A) is taken and the channel mixing 
is taken into account. 
Model results are denoted by lines. For $\eta'$ meson in panel (a), the experimental data  
~\cite{Csorgo} is shown  by the rectangle with the thin dotted 
vertical line $T=177$~MeV; see the text for the explanation. 
}
\label{scr_pole-realistic}
\end{figure}

Now we predict meson pole masses in the realistic case, 
taking the parameter set (A) and taking account of the channel mixing 
in model calculations. 
The results are shown for pseudoscalar mesons $\pi,K,\eta,\eta'$ 
in Fig. \ref{scr_pole-realistic}(a) and for 
scalar mesons $a_0,\kappa,\sigma,f_0$ in Fig. \ref{scr_pole-realistic}(b). 
For $\eta'$ meson, the pole mass in medium with finite $T$ 
was deduced from heavy-ion collision measurements 
as $M_{\eta'}^{\rm pole}(T)=340^{+375}_{-245}$~MeV~\cite{Csorgo}. 
In the analyses, $T=177$~MeV is taken as the default value and 
$T$ is varied systematically between 140 and 220 MeV. 
We then denote the experimental data~\cite{Csorgo} by 
the rectangle $(140~{\rm MeV} \le T \le 220~{\rm MeV}, 
95~{\rm MeV} \le M_{\eta'}^{\rm pole} \le 715~{\rm MeV})$ 
with the thin dotted vertical line standing 
for the default value $T=177$~MeV. 
The model result is  consistent with the experimental data. 
In general, $M_{\xi}^{\rm pole}$ is not smooth when the 
quark-pair production threshold is opened. 
This threshold effect is seen at $T=190$ MeV, e.g., for 
$\eta'$ meson.

As shown later in Sec. \ref{Discussions}, the channel mixing is negligible 
for $\eta'$ meson in $T > 160$~MeV, indicating that 
$\eta'$ meson is purely the ${\bar{s}s}$ state there. 
This result supports the ansatz in experimental analyses 
that $\eta'$ meson behaves just like a free particle in medium after 
it is produced.

%%%%%%%%%%%%%%%% Fig %%%%%%%%%%%%%%%%%%%%%
%%%%%%%%%%%%%

\subsection{Relation between pole and screening masses}
\label{Relation between pole and screening masses}

\begin{figure}[t]%[H]
\begin{center}
   \includegraphics[width=0.45\textwidth]{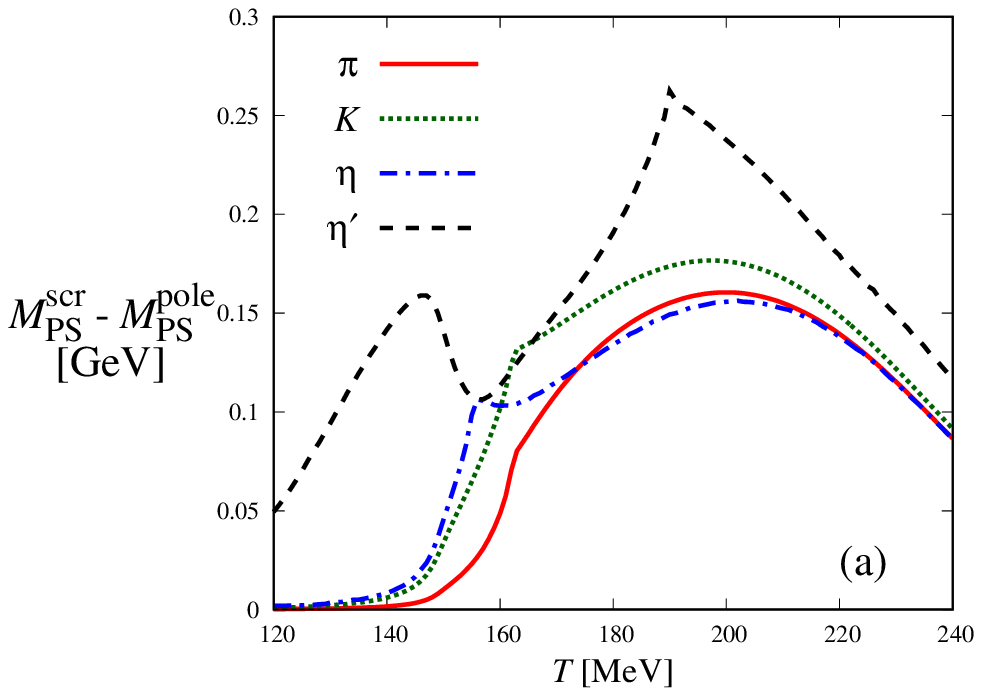}
   \includegraphics[width=0.45\textwidth]{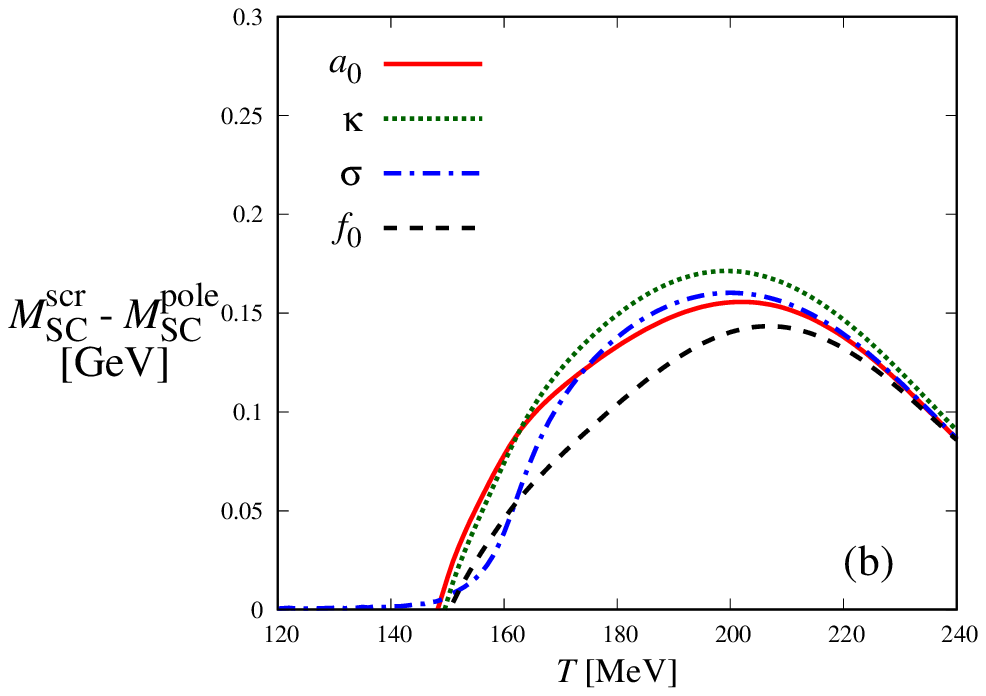}
\end{center}
 \caption{Difference between screening and pole masses 
for (a) pseudoscalar mesons 
$\pi,K,\eta,\eta'$ and (b) scalar mesons 
$a_0,\kappa,\sigma,f_0$. Model results are denoted by lines. 
In model calculations, the parameter set (A) is taken and the channel mixing 
 is taken into account. 
} 
\label{dif_pole_scr}
\end{figure}

Figure \ref{dif_pole_scr} shows $T$ dependence of the difference 
$M_{\xi}^{\rm scr}(T)-M_{\xi}^{\rm pole}(T)$ 
for (a) pseudoscalar mesons $\pi,K,\eta,\eta'$ and (b) scalar mesons 
$a_0,\kappa,\sigma,f_0$, where 
the parameter set (A) is taken and the channel mixing 
is taken into account in model calculations, 
whenever $T$ dependence of the difference is not smooth, 
it is due to the threshold effect. 
For pseudoscalar mesons, 
the difference tends to become larger for heavier meson. 
For scalar mesons, meanwhile, the difference is universal 
approximately: 
\bea
M_{\xi}^{\rm scr}(T)-M_{\xi}^{\rm pole}(T) \approx 
M_{\xi'}^{\rm scr}(T)-M_{\xi'}^{\rm pole}(T)
\label{mass-diff}
\eea
for $\xi \neq \xi'$. The deviation is about 35 MeV at $T\approx 200$ MeV. If $M_{\xi}^{\rm scr}(T)$, $M_{\xi'}^{\rm scr}(T)$ 
and $M_{\xi'}^{\rm pole}(T)$ are obtained with LQCD 
simulations, one can estimate $T$ dependence of 
$M_{\xi}^{\rm pole}(T)$ by using Eq. \eqref{mass-diff}.

%%%%%%%%%%%%%%%% Fig %%%%%%%%%%%%%%%%%%%%%
%%%%%%%%%%%%%

Next, the relation between 
$M_{\xi}^{\rm pole}(T)$ and $M_{\xi}^{\rm scr}(T)$ is considered 
through the ratios $M_{\xi}^{\rm pole}(T)/M_{\xi'}^{\rm pole}(T)$ and 
$M_{\xi}^{\rm scr}(T)/M_{\xi'}^{\rm scr}(T)$, 
where $\xi'$ is assumed to be a scalar (pseudoscalar) meson when 
$\xi$ is a scalar (pseudoscalar) meson. The identity 
\begin{equation}
 \frac{M_{\xi}^{\rm pole}(T)}{M_{\xi'}^{\rm pole}(T)}
  =
  \frac{M_{\xi}^{\rm scr}(T)}{M_{\xi'}^{\rm scr}(T)} 
\label{ratio-eq-0}
\end{equation}
is satisfied at both $T=0$ and $\infty$. The identity  at $T=0$ comes from the fact that 
${M_{\xi}^{\rm scr}(0)}={M_{\xi}^{\rm pole}(0)}$ for any meson. 
The identity at $T=\infty$ can be proven as follows. 
As mentioned in Sec. \ref{Meson screening masses}, 
in the large-$T$ limit all the 
$M_{\xi}^{\rm scr}(T)$ tend to $2\pi T$. 
Therefore, the ratio ${M_{\xi}^{\rm scr}}/{M_{\xi'}^{\rm scr}}$ becomes 1  
in the limit.
Similarly, the ratio 
${M_{\xi}^{\rm pole}}/{M_{\xi'}^{\rm pole}}$ approaches 1 
with respect to increasing $T$ as a consequence of the {\it effective} 
$SU(3)_{\rm V}$-symmetry restoration. The symmetry is broken 
by the fact $m_s \neq m_l$ in vacuum, 
but it is restored {\it effectively} at high $T$ because 
the symmetry breaking is the order of $(m_s - m_l)/T$ there; precisely speaking, 
for the flavor-singlet states, the symmetry is broken also by 
the quark-line disconnected diagrams, but the diagrams are suppressed 
by the Debye screening and the weakly interacting nature 
at high $T$ \cite{Bazavov:2012}.

\begin{figure}[t]%[H]
\begin{center}
   \includegraphics[width=0.45\textwidth]{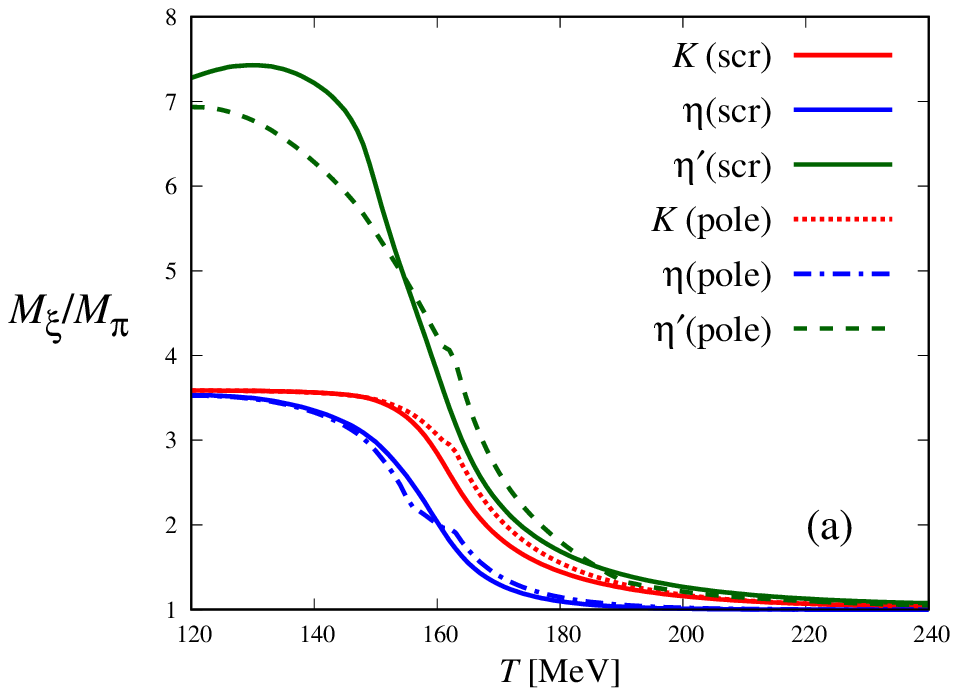}
   \includegraphics[width=0.45\textwidth]{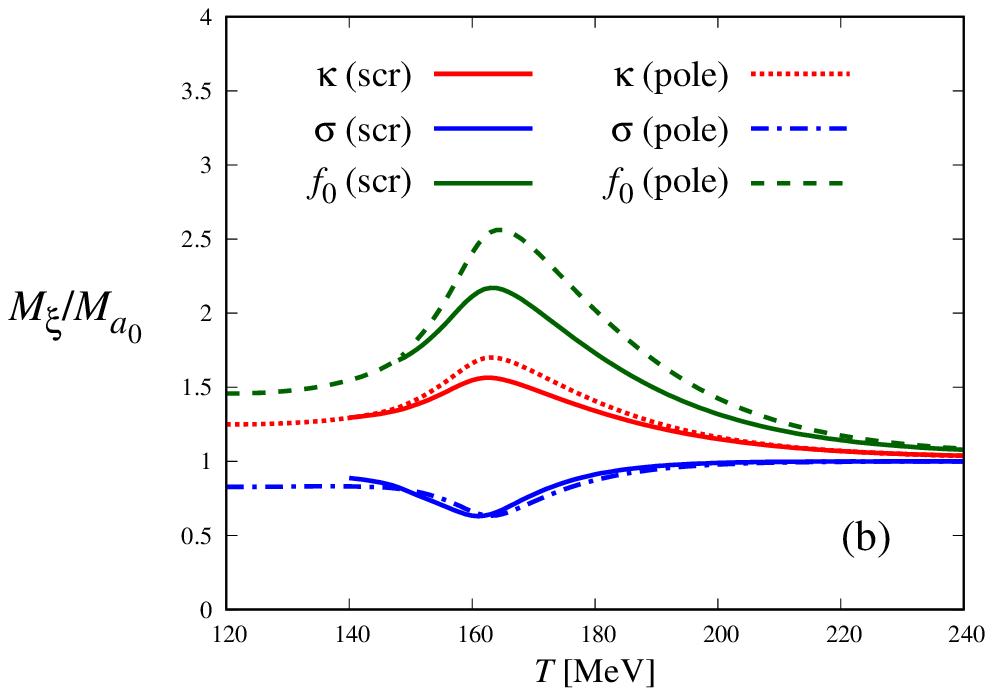}
\end{center}
 \caption{
$T$ dependence of 
${M_{\xi}^{\rm pole}}/{M_{\xi'}^{\rm pole}}$ and 
${M_{\xi}^{\rm scr}}/{M_{\xi'}^{\rm scr}}$ 
for 
(a) pseudoscalar mesons 
$(\xi=K,\eta,\eta',~\xi'=\pi)$ and (b) scalar mesons 
$(\xi=\kappa,\sigma,f_0,~\xi'=a_0)$. 
The ratios 
${M_{\xi}^{\rm scr}}/{M_{\xi'}^{\rm scr}}$ are denoted by 
solid lines, and the ratios 
${M_{\xi}^{\rm pole}}/{M_{\xi'}^{\rm pole}}$ are by 
dotted, dashed and dot-dashed lines. In model calculations, the parameter set (A) is taken and the channel mixing 
is taken into account. 
}
 \label{ratio-set-A}
\end{figure}

Figure \ref{ratio-set-A} shows the ratios as a function of $T$ 
for (a) pseudoscalar mesons ($\xi=K,\eta,\eta'$, $\xi'=\pi$) and (b) 
scalar mesons ($\xi=\kappa,\sigma,f_0$, $\xi'=a_0$). 
Qualitatively, the two ratios have similar $T$ dependence each other 
for both pseudoscalar and scalar mesons: Namely, 
\begin{equation}
 \frac{M_{\xi}^{\rm pole}(T)}{M_{\xi'}^{\rm pole}(T)}
  \simeq
  \frac{M_{\xi}^{\rm scr}(T)}{M_{\xi'}^{\rm scr}(T)} .
\label{ratio-eq}
\end{equation} 
Quantitatively, the relation \eqref{ratio-eq}
is well satisfied within 20\% error for pseudoscalar and scalar mesons. The relation is useful, because it allows us to 
estimate $M_{\xi}^{\rm pole}(T)$ 
for lighter $\xi$-meson  
from $M_{\xi'}^{\rm pole}(T)$ for heavier $\xi'$-meson  and 
$M_{\xi}^{\rm scr}(T)/M_{\xi'}^{\rm scr}(T)$ that may be obtainable with 
state-of-art LQCD simulations.

%%%%%%%%%%%%%%%% Fig %%%%%%%%%%%%%%%%%%%%%
%%%%%%%%%%%%%

\subsection{Discussion}
\label{Discussions}

\begin{figure}[t]%[H]
\begin{center}
   \includegraphics[width=0.45\textwidth]{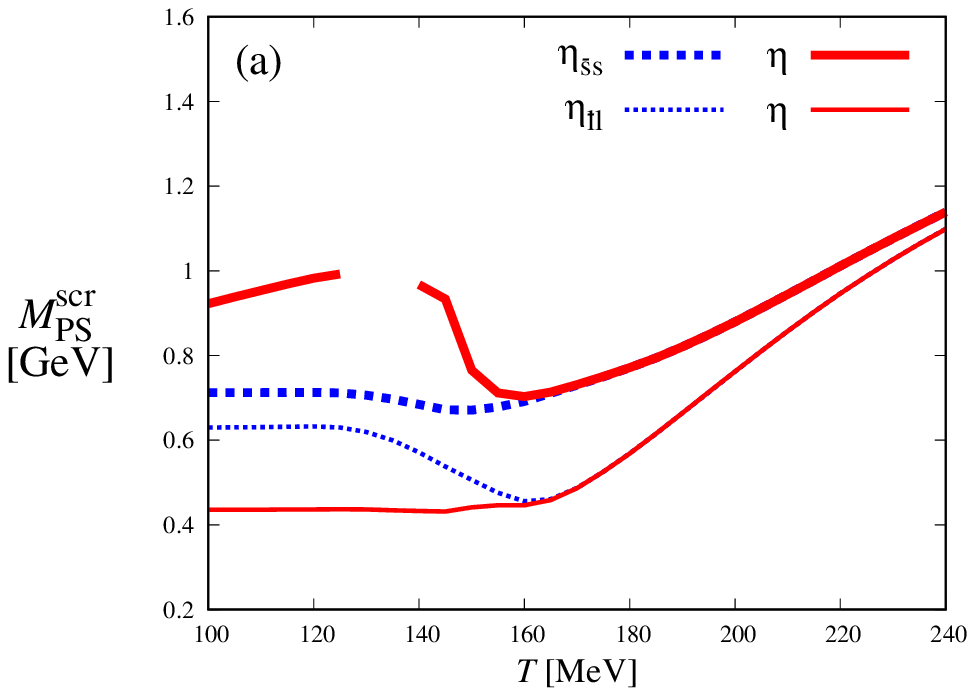}
   \includegraphics[width=0.45\textwidth]{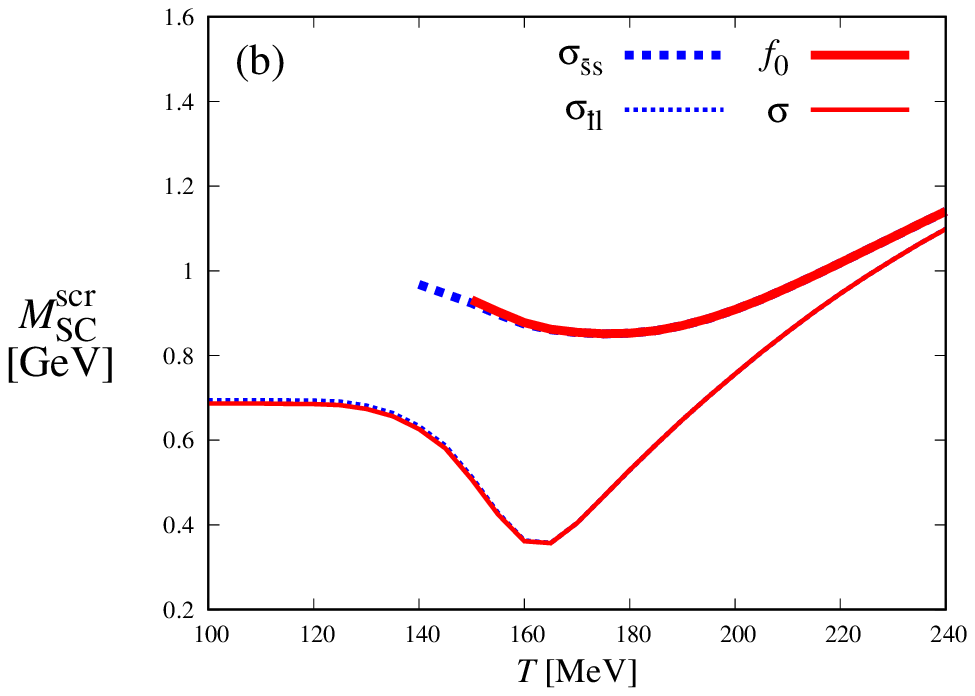}
\end{center}
 \caption{$T$ dependence of 
channel-mixing effects on (a) $\eta$- and $\eta'$-meson screening masses 
and (b) $\sigma$- and $f_0$-meson screening masses. 
In panel (a) (panel (b)), the thin and thick solid lines denote 
screening masses of $\eta$ and $\eta'$ ($\sigma$ and $f_0$)  mesons, 
respectively, and the thin and thick dashed lines correspond to 
screening masses of 
$\eta_{\bar{l}l}$ and $\eta_{\bar{s}s}$ ($\sigma_{\bar{l}l}$ and 
$\sigma_{\bar{s}s}$) channels, respectively. 
The parameter set (B) is taken in model calculations. 
}
\label{ch-mixing}
\end{figure}

$T$ dependence of the channel-mixing effect 
is investigated within model calculations. 
The parameter set (B) is taken in model calculations. 
In Fig.~\ref{ch-mixing}(a), 
the thin and thick solid lines denote 
the results of model calculations with the channel mixing 
for screening masses of $\eta$ and $\eta'$  mesons, 
respectively. 
Note that the lines are drawn when 
the condition $M_{\xi}^{\rm scr}(T)<M_{\rm th}$ is satisfied. 
When the channel mixing is switched off, 
the thin and thick solid lines are changed into 
the thin and thick dashed lines that stand for 
screening masses of $\eta_{\bar{l}l}$ and $\eta_{\bar{s}s}$ 
channels, respectively. As expected in Sec. \ref{results:Meson screening
masses}, the channel-mixing effect is large for 
$\eta$- and $\eta'$-meson screening masses in 
$T < 1.04T_c^\chi=160$~MeV. This is a result of the fact 
that the mass difference between the thin and thick dashed lines is small 
there; 
for example, the difference is 113 MeV at $T=140$ MeV. 
For $T >  1.04T_c^\chi=160$~MeV, the channel-mixing effect is negligible, 
since $G_{\eta_{\bar{s}s}\eta_{\bar{l}l}}  = 
 G_{\eta_{\bar{l}l}\eta_{\bar{s}s}} = 
G_{\rm D}(T)\sigma_l/\sqrt{2}$ is quite small 
in Eq. \eqref{gpi_eta} because of $\sigma_l \approx 0$.
In Fig.~\ref{ch-mixing}(b), 
the thin and thick solid lines stand for 
the results of model calculations with the channel mixing 
for screening masses of $\sigma$ and $f_0$ mesons, 
respectively, while 
the thin and thick dashed lines correspond to 
the results of model calculations without the channel mixing 
for screening masses of $\sigma_{\bar{l}l}$ and $\sigma_{\bar{s}s}$ channels, 
respectively. 
In the case of $\sigma$ and $f_0$ mesons, 
the channel-mixing effect is negligible for any $T$. 
This is because 
the mass difference between the thin and thick dashed lines is large 
in $T < 1.04 T_c^\chi=160$~MeV
(e.g., the difference is 335 MeV at $T=140$ MeV) 
and 
$G_{\sigma_{\bar{s}s}\sigma_{\bar{l}l}}  = 
 G_{\sigma_{\bar{l}l}\sigma_{\bar{s}s}} = 
- G_{\rm D}(T)\sigma_l/\sqrt{2} \approx 0$ 
because of $\sigma_l \approx 0$ in $T >  1.04 T_c^\chi=160$~MeV. 
Consequently, the channel mixing as the characteristics of 
the disconnected diagrams is important only 
for $\eta$- and $\eta'$-meson screening masses 
in $T < 1.04 T_c^\chi=160$~MeV. Also for $\eta$- and $\eta'$-meson pole masses, we can take  
the same conclusion.

%%%%%%%%%%%%%%%% Fig %%%%%%%%%%%%%%%%%%%%%
%%%%%%%%%%%%%

%%%%%%%%%%%%%%%%%%%%%%%%%%%%%%%%%%%%%%%%%%%%%%%%%%%%%%%%%%%%%%%%%%%%%%%%%%%
%%%%% Summary
%%%%%%%%%%%%%%%%%%%%%%%%%%%%%%%%%%%%%%%%%%%%%%%%%%%%%%%%%%%%%%%%%%%%%%%%%%%
\section{Summary}
\label{Summary}

We have proposed a practical effective model by introducing 
$T$-dependent coupling strengths, $G_{\rm S}(T)$ and $G_{\rm D}(T)$, 
to four-quark and six-quark KMT  interactions in the 2+1 flavor PNJL model. 
$T$ dependence of $G_{\rm D}(T)$ is determined from 
LQCD data~\cite{Cheng:2010fe} on 
$\Delta M_{a_0,\pi}^{\rm scr}(T)$ 
in $T > 1.1 T_c^\chi=170$~MeV 
where only the $U(1)_{\rm A}$-symmetry breaking survives. 
Similarly, $T$ dependence of $G_{\rm S}(T)$ is determined 
from LQCD data associated with the chiral-symmetry restoration, i.e., the
renormalized chiral 
condensate~$\Delta_{l,s}(T)$~\cite{Cheng:2008} and 
the pseudocritical temperature $T_c^{\chi}=154\pm 9$ MeV~\cite{Borsanyi:2010bp,Bazavov:2011nk} of chiral transition. 
In LQCD simulations of Ref.~\cite{Cheng:2008,Cheng:2010fe}, 
the lattice setting is that 
$m_l/m_s=1/10$ and $M_{\pi}^{\rm pole}(0)=176$ MeV. 
In model calculations, 
$m_l$ and $m_s$ are slightly changed from the realistic parameter set (A) 
of $M_{\pi}^{\rm pole}(0)=138$ MeV and $M_{K}^{\rm pole}(0)=495$ MeV so as to 
become consistent with the lattice setting. 
This parameter set is referred to as set (B).

The present model with the parameter set (B) reproduces 
LQCD data~\cite{Cheng:2010fe} on 
$M_{\xi}^{\rm scr}(T)$ for both pseudoscalar mesons 
$\pi,K,\eta_{\bar{s}s}$ and scalar mesons 
$a_0,\kappa,\sigma_{\bar{s}s}$ in $T > 1.04T_c^{\chi}=160$ MeV.  
Meanwhile, in $T < 1.04T_c^{\chi}=160$ MeV, the 
agreement between model results and LQCD data is good for 
pseudoscalar $\pi,K$ mesons and pretty good for scalar 
$a_0,\kappa,\sigma_{\bar{s}s}$ mesons. 
For $\eta_{\bar{s}s}$ meson, 
the model result overestimates 
LQCD data by about $10\% \sim 30\%$ 
in $T < 1.04T_c^{\chi}=160$ MeV, but the deviation 
decreases rapidly as $T$ increases from $160$ MeV.

In finite-$T$ LQCD simulations of Ref.~\cite{Cheng:2010fe}, 
the disconnected diagrams are neglected when $M_{\eta_{\bar{s}s}}^{\rm scr}(T)$ is calculated. 
As a consequence of this approximation, the $\eta_{\bar{s}s}$ channel 
is decoupled with the $\eta_{\bar{l}l}$ channel. 
We then divide the disconnected-diagram effects into 
the channel-mixing effect and 
the remaining disconnected-diagram effects acting on the diagonal elements of 
the correlation-function matrix ${\boldsymbol \chi}_{\xi}$. 
The model calculation includes 
the remaining disconnected-diagram effects implicitly, even if the channel 
mixing is switched off. The deviation between model calculations and 
LQCD data for $\eta_{\bar{s}s}$ meson in $T < 1.04T_c^{\chi}=160$ MeV may 
stem from the remaining disconnected-diagram effects implicitly 
included in model calculations.

Using this practical effective model with the realistic parameter set (A), 
we have predicted meson pole masses $M_{\xi}^{\rm pole}(T)$ for 
pseudoscalar mesons $\pi,K,\eta,\eta'$ and scalar mesons 
$a_0,\kappa,\sigma,f_0$.  
This prediction makes it possible to compare the $M_{\xi}^{\rm pole}(T)$ 
evaluated from LQCD data with the corresponding experimental results directly. 
In fact, we have shown that for $\eta'$ meson the model prediction  is 
consistent with the experimental value~\cite{Csorgo} at finite $T$ measured in heavy-ion collisions. 
If experimental data are not available for 
$M_{\xi}^{\rm pole}(T)$ of interest, such a prediction may be helpful in 
experimental analyses.

We have shown that the relation 
$M_{\xi}^{\rm scr}(T)-M_{\xi}^{\rm pole}(T) \approx 
M_{\xi'}^{\rm scr}(T)-M_{\xi'}^{\rm pole}(T)$ 
is pretty good when $\xi$ and $\xi'$ are scalar mesons, 
and have pointed out that the relation 
$M_{\xi}^{\rm scr}(T)/M_{\xi'}^{\rm scr}(T) \approx 
M_{\xi}^{\rm pole}(T)/M_{\xi'}^{\rm pole}(T)$ is 
well satisfied within 20\% error when $\xi$ 
and $\xi'$ are pseudoscalar mesons and also when $\xi$ 
and $\xi'$ are scalar mesons. 
These relations may be useful to estimate $M_{\xi}^{\rm pole}$ 
for lighter $\xi$-meson  
from $M_{\xi'}^{\rm pole}$ and $M_{\xi'}^{\rm scr}$ for heavier $\xi'$-meson  
that may be obtainable with 
state-of-art LQCD simulations.

We have also found 
that the channel mixing as the characteristics of 
the disconnected diagrams is important only 
for $\eta$- and $\eta'$-meson masses 
in $T < 1.04 T_c^\chi=160$~MeV. 
This indicates that 
$\eta'$ meson is purely the ${\bar{s}s}$ state 
in $T \ga 1.04 T_c^\chi=160$~MeV. 
This result supports the ansatz in experimental analyses 
that $\eta'$ meson behaves just like a free particle 
in medium after it is produced.

\noindent
\begin{acknowledgments}
M. I, H. K., and M. Y. are supported by Grants-in-Aid for Scientific Research (
No. 27-3944, No. 26400279 and No. 26400278) from the Japan Society for
 the Promotion of Science (JSPS). The authors thank to J. Takahashi,
 Y. Maezawa and M. Oka for useful comments.
\end{acknowledgments}

%%%%%%%%%%%%%%%%%%%%%%%%%%%%%%%%%%%%%%%%%%%%%%%%%%%%%%%%%%%%%%%%%%%%%%%%%%%%%%%
%%%%% Appendix
%%%%%%%%%%%%%%%%%%%%%%%%%%%%%%%%%%%%%%%%%%%%%%%%%%%%%%%%%%%%%%%%%%%%%%%%%%%%%%%%
\noindent
\appendix
\section{Difficulty of meson pole-mass calculations \\ 
in LQCD simulations at finite $T$}
\label{Difficulty of meson pole mass calculations in LQCD}

In general, the meson pole mass $M_{\xi}^{\rm pole}$ and its decay width 
$\Gamma_{\xi}$ are determined from 
energy ($\omega$) dependence of the spectral function 
$\rho_{\xi\xi}(\omega,{\bf p},T)$, 
where the spatial momentum ${\bf p}$ is set to zero when we calculate 
the pole mass. 
In the Matsubara formalism, $\rho_{\xi\xi}(\omega,{\bf p},T)$ is 
related to the mesonic 
correlation function $\mathcal{G}_{\xi\xi}(\tau,{\bf p},T)$ as 
\begin{eqnarray}
\mathcal{G}_{\xi\xi} (\tau,{\bf p},T) &=& \int_0^{\infty} d\omega
 ~\rho_{\xi\xi}(\omega,{\bf p},T)K(\omega,\tau,T),
 \label{spectral}
 \\
K(\omega,\tau,T) &=& \frac{\cosh{\left[\omega(\tau - 1/2T)\right]}}{\sinh{(\omega/2T)}},
\end{eqnarray}
where $\tau$  represents imaginary time satisfying $0\le \tau \le 1/T$. 
If the $\mathcal{G}_{\xi\xi}(\tau,{\bf p},T)$ is known, 
the spectral function $\rho_{\xi\xi}(\omega,{\bf p},T)$ 
can be obtained from the $\mathcal{G}_{\xi\xi}$ by solving 
the integral equation \eqref{spectral} for $\rho_{\xi\xi}$. 
In LQCD simulations at finite $T$, 
the $\mathcal{G}_{\xi\xi}(\tau,{\bf p},T)$ is obtained only for a limited number of 
 $\tau$  up to $1/T$. 
In general, the number is too small to construct $\rho_{\xi\xi}$ 
as a continuous function of $\omega$.

%%%%%%%%%%%%%%%%%%%%%%%%%%%%%%%%%%%%%%%%%%%%%%%%%%%%%%%%%%%%%%%%%%%%%%%%%%%%%%%%%%%%% 
% References 
%%%%%%%%%%%%%%%%%%%%%%%%%%%%%%%%%%%%%%%%%%%%%%%%%%%%%%%%%%%%%%%%%%%%%%%%%%%%%%%%

\end{document}